\documentclass[usenatbib]{mn2e}    
\usepackage[final]{graphicx}
\usepackage{epstopdf}
\usepackage{mathptmx}
\usepackage{float}
\usepackage{fixltx2e}
\usepackage{natbib}
\usepackage{amsmath}
\usepackage{amssymb}
\usepackage{color}
\usepackage{appendix}
\usepackage[mathscr]{euscript}

\newcommand{\mbound}{M_{\text{bound}}}

\newcommand{\gimic}{\textsc{gimic}}
\newcommand{\msun}{\mathrm{M}_\odot}
\newcommand{\mstar}{M_\text{star}}

\title[Environmental star formation quenching]{Star formation quenching in simulated group and cluster galaxies: When, how, and why?}
\author[Y.~Bah\'{e} and I.~McCarthy]{Yannick~M.~Bah\'{e}\thanks{ybahe@mpa-garching.mpg.de}$^{1,2}$ and Ian G.~McCarthy$^{3}$\\
$^1$ Max-Planck-Institut f\"{u}r Astrophysik, Karl-Schwarzschild Str. 1, 85748 Garching, Germany\\
$^2$ Institute of Astronomy, University of Cambridge, Madingley Road, Cambridge CB3 0HA, United Kingdom\\
$^3$ Astrophysics Research Institute, Liverpool John Moores University, IC2, 146 Brownlow Hill, Liverpool L3 5RF, United Kingdom\\
}

\begin{document}
\label{firstpage}
\maketitle

\begin{abstract}
Star formation is observed to be suppressed in group and cluster galaxies compared to the field. To gain insight into the quenching process, we have analysed $\sim 2000$ galaxies formed in the \gimic{} suite of cosmological hydrodynamical simulations. The time of quenching varies from $\sim 2$ Gyr before accretion (first crossing of $r_{200,\,c}$) to $> 4$ Gyr after, depending on satellite and host mass. Once begun, quenching is rapid ($\lesssim 500$ Myr) in low-mass galaxies ($\mstar < 10^{10}\, \msun$), but significantly more protracted for more massive satellites. The simulations predict a substantial role of outflows driven by ram pressure -- but not tidal forces -- in removing the star-forming interstellar matter (ISM) from satellite galaxies, especially dwarfs ($\mstar \approx 10^9\, \msun$) where they account for nearly two thirds of ISM loss in both groups and clusters. Immediately before quenching is complete, this fraction rises to $\sim 80$ per cent even for Milky Way analogues ($\mstar \approx 10^{10.5}\, \msun$) in groups ($M_\text{host} \approx 10^{13.5}\, \msun$). We show that (i) ISM stripping was significantly more effective at early times than at $z = 0$; (ii) approximately half the gas is stripped from `galactic fountains' and half directly from the star forming disk; (iii) galaxies undergoing stripping experience ram pressure up to $\sim 100$ times the average at a given group/cluster-centric radius, because they are preferentially located in overdense ICM regions. Remarkably, stripping causes at most half the loss of the extended gas haloes surrounding our simulated satellites. These results contrast sharply with the current picture of strangulation -- removal of the ISM through star formation after stripping of the hot halo -- being the dominant mechanism quenching group and cluster satellites.
\end{abstract}

\begin{keywords}
galaxies: clusters: general --- galaxies: evolution --- galaxies: haloes --- galaxies: interactions --- galaxies: intergalactic medium --- galaxies: ISM
\end{keywords}


\section{Introduction}
\label{sec:introduction}

The internal properties of galaxies in groups and clusters are strongly correlated with their local environment. Isolated `field' galaxies like our Milky Way are typically `late-type' with a spiral disc containing young blue stars, whereas groups and clusters are dominated by `early-type' galaxies, apparently structureless ellipticals mostly composed of old, red stars (for recent reviews see e.g.~\citealt{Boselli_Gavazzi_2006, Blanton_Moustakas_2009}). On a physical level, this `colour--density relation' corresponds to reduced star formation in group and cluster galaxies (e.g.~\citealt{Balogh_et_al_1999,Lewis_et_al_2002,Gomez_et_al_2003,Kauffmann_et_al_2004,Weinmann_et_al_2006,Peng_et_al_2010,Peng_et_al_2012,Wetzel_et_al_2012}), which is in turn believed to be causally linked to a lack of cold star forming gas (e.g.~\citealt{Haynes_Giovanelli_1984,Elmegreen_et_al_2002,Gavazzi_et_al_2006,Fabello_et_al_2012, Catinella_et_al_2013, Hess_Wilcots_2013}). However, there is currently no clear consensus on the cause of this gas depletion. This constitutes an important gap in our understanding of galaxy formation and evolution: $\sim 30$ per cent of all galaxies reside in groups (e.g.~\citealt{Budzynski_et_al_2012}), and clusters in particular are also of interest as precision probes of cosmology, which requires an accurate understanding of their internal workings (e.g.~\citealt{Voit_2005,Allen_et_al_2011,Kravtsov_Borgani_2012}).
 
Within the hierarchical $\Lambda$CDM structure formation paradigm, galaxies are accreted onto groups and clusters from the field, so that the observed trends hint at some form of transformation during infall. A large number of mechanisms have been proposed that could cause such transformations. The motion of a galaxy relative to the intragroup-/cluster medium (ICM\footnote{We will refer to this extended gaseous halo as `ICM' in the context of both groups and clusters.}) causes a force on its gas through ram pressure \citep{Gunn_Gott_1972} or momentum transfer caused by the (presently poorly constrained) ICM viscosity \citep{Nulsen_1982}. If strong enough, this force can strip star forming gas from the galaxy. Tidal forces due to the group or cluster potential can have a similar effect (e.g. \citealt{Willman_et_al_2004}), as can heat transfer from the ICM to the galaxy which may lead to `evaporation' of its gas \citep{Cowie_Songaila_1977}, or `galaxy harassment' \citep{Moore_et_al_1996, Moore_et_al_1999}. 

More indirectly, it is also possible to quench star formation by only throttling the supply of \emph{new} cold, star-forming gas through removal of the hot gas haloes or cold accretion flows which are thought to fulfil this role in field galaxies (\citealp{White_Rees_1978, White_Frenk_1991}). Star formation will then cease after some delay in which the remaining cold gas is used up, which is why this process is often referred to as `starvation' or `strangulation' (\citealt{Larson_et_al_1980}; see also \citealt{Balogh_et_al_2000, McCarthy_et_al_2008b}). This normal consumption removes not just gas turned into stars, but also that which is expelled through winds driven by supernovae and/or accreting supermassive black holes, so that the gas consumption time may well be significantly shorter than inferred from a simple comparison between the star-formation rate and available gas mass (see also \citealt{McGee_et_al_2014}). Furthermore, the consumption of cold gas may be sped up if star formation is temporarily enhanced, for example through the compressive effects of ram pressure \citep{Tonnesen_Bryan_2012} or tidal forces \citep{Byrd_Valtonen_1990}.

All these mechanisms are at least plausible, and observations of individual galaxies have provided direct evidence for the action of some of them. For example, some galaxies in nearby clusters such as Virgo exhibit long tails of H\textsc{i} emitting gas that were likely stripped by ram pressure \citep{Gavazzi_Jaffe_1987, Gavazzi_et_al_1995, Kenney_et_al_2004, Chung_et_al_2007, Chung_et_al_2009, Abramson_et_al_2011, Merluzzi_et_al_2013}, some of which show signs of star formation in dense clumps (e.g.~\citealt{Ebeling_et_al_2014, Roediger_et_al_2014}). X-ray emitting tails behind cluster galaxies such as ESO 137-009 \citep{Sun_et_al_2006} are strong evidence for ram pressure stripping of hot gas, leading to strangulation. 

The question remains, however, which of these mechanisms are actually important in driving the environmental trends that we observe in the galaxy population as a whole. For example, slow processes such as strangulation can only be effective if they are not interrupted by more violent ones such as tidal or ram pressure stripping, which may or may not be the case. 

Answering this question requires large, representative samples of galaxies, such as from the \emph{Sloan Digital Sky Survey} (SDSS; \citealt{York_et_al_2000}) or \emph{Galaxy and Mass Assembly} survey (GAMA; \citealt{Baldry_et_al_2010}). These data have shown that low-redshift galaxies in all environments exhibit a strong bimodality in colour and sSFR, with an `active' (blue) and `passive' (red) peak separated by a relative dearth of galaxies in the intervening `green valley' (e.g.~\citealt{Strateva_et_al_2001, Baldry_et_al_2004, Peng_et_al_2010, Wetzel_et_al_2012}). Moreover, a number of authors have shown that only the relative proportion of galaxies in either peak varies with environment (from the scale of poor groups), but \emph{not} the location of either the blue peak or the green valley, nor the depth of the latter (e.g.~\citealt{Balogh_et_al_2004b, Wetzel_et_al_2012}). 

These results suggest that the local environment must influence galaxies even at mass-scales well below those of massive galaxy clusters (see also \citealt{Balogh_McGee_2010}). Any environmental transformation cannot be an \emph{immediate} consequence of accretion onto a group or cluster, because some galaxies are observed to form stars even in massive clusters, but when quenching occurs, it must be rapid to explain the absence of both a shift in the blue peak and an increased population of galaxies in the green valley \citep{Wetzel_et_al_2012}.
Together, these clues have been argued to support a ``delayed-then-rapid'' quenching scenario \citep{Wetzel_et_al_2013}.  

There is, however, a substantial and growing body of observational evidence in conflict with this interpretation. For example, several studies have presented evidence in favour of a slowly declining star formation rate of group and cluster galaxies (e.g. \citealt{von_der_Linden_et_al_2010, Vulcani_et_al_2010, de_Lucia_et_al_2012, Rasmussen_et_al_2012, Taranu_et_al_2014}), which would favour more gradual, slow transformation mechanisms. \citet{Fabello_et_al_2012}, on the other hand, compared H\textsc{i} observations to predictions from models invoking the strangulation mechanisms and found evidence for the wide-spread action of (rapid) ram pressure stripping, even in low-mass haloes of mass $\sim 10^{13}\, \msun$ (see also \citealt{Boesch_et_al_2013, Catinella_et_al_2013}). Additionally, increasing evidence points to galaxies being affected well beyond the `virial' radius, the traditionally assumed boundary of the influence of groups and clusters (\citealt{Balogh_et_al_1999, von_der_Linden_et_al_2010, Lu_et_al_2012, Behroozi_et_al_2014}; see also \citealt{Bahe_et_al_2013}). Finally, while the colour--density relation itself has been shown to extend at least to redshift $z \approx 2$ (e.g.~\citealt{Cooper_et_al_2006,Cooper_et_al_2008}), individual features such as the sparsity of the green valley appear to be much less pronounced at higher redshift \citep{Balogh_et_al_2011, Mok_et_al_2013}. 

One conclusion that seems to follow from this complex and seemingly contradictory observational evidence is that the evolution of group and cluster galaxies is not a simple process, and determining a `typical' sequence of events that all galaxies have experienced may well prove elusive. It is, for example, well possible that different processes dominate the transformation of galaxies of different mass or type, in differently massive hosts, and at different epochs, introducing the possibility of subtle selection effects when comparing observations and theory. Even within an individual galaxy, different processes may drive quenching in the inner and outer regions, which could give rise to selection effects due to varying observational apertures: spectra from the SDSS, for instance, cover only the central 3'' of a galaxy whereas gas observed to be stripped into H$\textsc{i}$ emitting tails may originate from well outside this zone (but see \citealt{McBride_McCourt_2014}). 

Simple toy models compatible with observational constraints (e.g.~\citealt{Wetzel_et_al_2013, Mok_et_al_2014}), or even more sophisticated semi-analytic models (e.g.~\citealt{Guo_et_al_2011}) may therefore be fundamentally limited in their power to answer the question of which mechanism dominates star-formation quenching under which conditions, which motivates a more self-consistent theoretical approach based on hydrodynamical simulations. Although inevitably only an imperfect model of galaxies in the real Universe, these simulations can nevertheless be an invaluable tool, because the availability of large amounts of observationally inaccessible information (such as full 6D phase-space coordinates and the evolution of individual galaxies over cosmic time) means that they can be studied in exquisite detail. Insight gained in this way can then be applied to real galaxies as well.

Such simulations have already been used to study either idealised model galaxies or the evolution of individual groups and clusters. \citet{Abadi_et_al_1999}, for example, exposed individual model galaxies to winds typical of those experienced by group and cluster galaxies, and concluded that ram pressure was only of relevance to galaxies in the cores of massive clusters, but not in the less dense environment of galaxy groups (see also \citealt{Gunn_Gott_1972,Brueggen_DeLucia_2008}). \citet{Tonnesen_et_al_2007}, on the other hand, simulated a galaxy cluster with $\sim 100$ member galaxies from cosmological initial conditions and found evidence for a significant effect of ram pressure. This was shown to be in part due to a large variation of ram pressure values at the same cluster-centric radius, so that some galaxies can be subject to much stronger forces than what would be estimated from simple models assuming spherical symmetry \citep{Tonnesen_Bryan_2008}. \citet{Feldmann_et_al_2011} modelled the evolution of a galaxy group with around a dozen members and found that mergers were the key driver producing elliptical galaxies, together with a stop of cosmological gas accretion in satellites.

While these studies have already shed some further light on the transformation mechanisms operating under certain conditions, determining which mechanisms are overall important in affecting the group and cluster galaxy population requires a detailed analysis of large simulations with a wide range of galaxy and group/cluster masses. This is the objective of the present study, where we aim to build on the above-mentioned works to investigate star-formation quenching in a large sample of $\sim 2000$ galaxies formed from cosmological initial conditions, spanning two decades in stellar and host mass, from the cosmological hydrodynamical simulation suite \gimic{} \citep{Crain_et_al_2009}. The zoomed initial conditions of these simulations (see Section 2) result in an extreme dynamical range of $\sim$ 6 orders of magnitude, resolving small galaxies of $\mstar \approx 10^9\, \msun$ while at the same time modelling a massive galaxy cluster of $M > 10^{15}\, \msun$. In addition, the simulations include a large number of state-of-the art astrophysics modules, which allows us to investigate the interplay between these effects (such as star-formation and supernova feedback) with environmental factors such as ram pressure stripping. \gimic{} \emph{field} galaxies have been shown to reproduce a large number of observational trends, including the L$_\text{X}$-L$_\text{K}$ scaling relation \citep{Crain_et_al_2010a}, satellite and stellar haloes \citep{Font_et_al_2011, McCarthy_et_al_2012a}, as well as scaling relations between stellar mass and rotation velocity, halo mass and galaxy size \citep{McCarthy_et_al_2012b}. Although the simulations are not completely realistic, they should therefore still be well-suited to investigating the transformations occurring when galaxies are accreted by a group or cluster.

This paper is structured as follows. In Section 2, we give a brief overview of the simulations and describe how we select galaxies, groups and clusters. Section 3 shows the extent of star formation quenching in \textsc{gimic}, followed by an analysis of the mechanisms responsible for removing ISM gas from satellite galaxies in Section 4. In Section 5 we investigate in detail the environmentally driven expulsion of star-forming gas. The reasons for the lack of ISM replenishment in group and cluster galaxies are analysed in Section 6, before we summarise and discuss our conclusions in Section 7. All masses and distances are given in physical units unless otherwise specified. A flat $\Lambda$CDM cosmology with Hubble parameter $h =$ H$_{0}/(100\,{\rm km}\,{\rm s}^{-1}{\rm Mpc}^{-1}) = 0.73$, dark energy density parameter $\Omega_\Lambda = 0.75$ (dark energy equation of state parameter $w=-1$), and matter density parameter $\Omega_{\rm M} = 0.25$ is used throughout this paper. Unless specified otherwise, halo radii are calculated as $r_{200} = r_{200, c}$, the radius within which the mean density equals 200 times the critical density of the Universe.


\section{Simulations and Analysis}
\label{sec:simulations}
\label{sec:gimic}

\subsection{The \gimic{} simulations}
Running cosmological hydrodynamical simulations of large volumes with high resolution --- a requirement to simultaneously include rare systems such as massive galaxy clusters and resolve individual galaxies ---  is currently prohibitively expensive. A promising way forward is therefore the use of `zoomed initial conditions' (e.g.~\citealt{Katz_et_al_1994, Navarro_et_al_2004}), in which only a small, carefully chosen part of a large cosmological volume is simulated at full resolution with hydrodynamics. Our analysis is based on a set of such zoomed simulations, the \textsc{Galaxies -- Intergalactic Medium Interaction Calculations} (\gimic{}; \citealt{Crain_et_al_2009}; see also \citealt{Schaye_et_al_2010}). We refer the reader to these papers for a full discussion of the simulations, and only summarise their main features relevant to this study.

The \textsc{gimic} simulations are a set of five re-simulations of nearly spherical regions with radius $\sim 20\, h^{-1}$ Mpc (comoving) of varying mean density, extracted from the Millennium Simulation \citep{Springel_et_al_2005}. The regions are chosen so that at $z = 1.5$ their average densities differ from the cosmic mean by (-2, -1, 0, +1, +2) $\sigma$, where $\sigma$ is the rms mass fluctuation on a scale of $18\, h^{-1}$ Mpc at this redshift. \textsc{gimic} therefore includes many groups and clusters of galaxies, including a particularly massive one with $\log_{10}$ ($M_{200} / \msun$) $\approx 15.2$ at $z=0$, at the centre of the +2$\sigma$ sphere. We note that the Coma cluster is only slightly more massive than this, with $\log_{10}$ ($M_{200} / \msun$) $\approx 15.4$ \citep{Kubo_et_al_2007}.

The simulations were carried out at 3 different resolutions: `low', `intermediate', and `high'. The `low' resolution is the same as in the original Millennium Simulation while the `intermediate' and `high' versions have $8$ and $64$ times better mass resolution, respectively.  As only the $-2\sigma$ and $0\sigma$ regions have been run at high resolution (owing to prohibitive computational expense), we use the intermediate-resolution simulations here.  These have a baryon particle mass of $m_\text{gas} \approx 1.16 \times 10^7 h^{-1} \msun$ with a gravitational softening that is 1 $h^{-1}$ kpc in physical space at $z \le 3$ and is fixed in comoving space at higher redshifts.  Thus, even relatively low-mass galaxies ($\mstar \approx 10^9\, \msun$) are resolved into several hundred particles\footnote{\citet{McCarthy_et_al_2008b} demonstrated that the stripped fractions are numerically converged at this number of resolution elements. For an example image of a \gimic{} disk galaxy --- albeit at the `high' resolution level --- see Fig.~1 of \citet{Font_et_al_2011}}.

The simulations were carried out with the TreePM-SPH code \textsc{Gadget-3} (last described in \citealt{Springel_2005}) and include significantly modified prescriptions for star formation \citep{Schaye_DallaVecchia_2008}, metal-dependent radiative cooling in the presence of a \citet{Haardt_Madau_2001} UV/X-ray background \citep{Wiersma_et_al_2009a}, as well as stellar evolution and chemodynamics \citep{Wiersma_et_al_2009b}. The effect of feedback from core collapse supernovae is modelled according to \citet{DallaVecchia_Schaye_2008}. In essence, this model prescribes that shortly after a gas particle is turned into a star particle, a small number (on average four) of its neighbouring gas particles receive a velocity `kick' of $v_\text{kick} = 600\, \text{km s}^{-1}$ in a random direction. As discussed in \citet{DallaVecchia_Schaye_2008}, the energy injected into the ISM in this way --- which corresponds to around 80 per cent of the available supernova energy assuming a \citet{Chabrier_2003} IMF --- can drive gas outflows similar to those observed in the real Universe. 

However, the \gimic{} simulations do not include a prescription for feedback due to accreting supermassive black holes (`AGN feedback'). This is probably the reason why massive galaxies with $\mstar \gtrsim 10^{11}\, \msun$ in \textsc{gimic} suffer from `over-cooling' and have artificially high gas densities in their central regions (see \citealt{Crain_et_al_2009} and \citealt{McCarthy_et_al_2012b}).  As a result, we do not expect realistic predictions for massive galaxies and therefore limit our analysis to the range $\log_{10} (\mstar / \msun) = [9.0 - 11.0]$. We discuss the implications of these shortcomings further in Section \ref{sec:discussion}.

\subsection{Identifying and tracing galaxies and groups/clusters}
\label{sec:tracing}
The result of the simulations is output at several dozen snapshots\footnote{The exact redshifts of these snapshots vary between the individual simulations, but the average time gap between them is $\sim 250$ Myr.} between redshifts $z = 10$ and $z = 0$. Within these, Friends-of-Friends (FoF) haloes and the gravitationally self-bound subhaloes therein are identified by the \textsc{subfind} algorithm (\citealt{Dolag_et_al_2009}; see also \citealt{Springel_et_al_2001b}). We then trace these subhaloes between snapshots using the unique IDs of their constituent stellar and dark matter particles, as described in detail in Appendix \ref{sec:app_tracing}. In essence, we find for each subhalo in one output snapshot the one in the next snapshot sharing most of its dark matter and stellar mass, accounting for the possibility of mergers, formation of new subhaloes, and temporary non-identification of existing ones (e.g.~\citealt{Muldrew_et_al_2011}). 

Galaxy groups and clusters are identified at redshift $z = 0$ as FoF haloes with a total mass $\mbound \geq 10^{13.0}\, \msun$, where $\mbound$ is the total mass of particles gravitationally bound to the halo (see \citealt{Bahe_et_al_2013}). Their progenitors at higher redshift are taken as the FoF haloes containing the progenitors of their main (central) subhalo as determined by our tracing procedure. As \emph{galaxies}\footnote{In the following, we shall refer to the physical entity of a galaxy as such, i.e.~the system which is identified at many different points in time. Each of these occurrences will be referred to as a galaxy `observation' or `snapshot'.} we select those subhaloes which have a (gravitationally bound) stellar mass of $\mstar \geq 10^9 \msun$ in at least one snapshot, and are identified in at least five snapshots. The second cut eliminates `galaxy' identifications which are actually transient (but self-bound) substructures in more massive galaxies such as parts of spiral arms (see \citealt{Bahe_et_al_2012}) and are typically only identified in at most a few consecutive snapshots before ``merging'' with their host galaxy. 

One important subtlety is that by temporarily `cutting off' parts of its star-forming spiral arms, a galaxy can  \emph{appear} to suddenly become gas-poor and quenched, without any physical quenching taking place at all. To avoid this, we treat particles in such sub-subhaloes as belonging to the (main) galaxy subhalo, provided they have been directly assigned to it by \textsc{subfind} in at least one previous snapshot. This ensures that we still separate off particles in sub-subhaloes that have not (yet) been part of the galaxy, which is important to distinguish, for example, star formation in infalling satellites from lack thereof in a central galaxy.  

Overall, these cuts leave us with $\sim 14,200$ reliably traced galaxies which form the basis of our study. From these, we extract two samples: One set of 3580 `satellite' galaxies, which at redshift $z = 0$ are located within $5\, r_{200}$ from one of our selected groups and clusters\footnote{We define $r_{200}$ as $r_{200,\,c}$, the radius inside which the mean density equals 200 times the \emph{critical} density of the Universe. Note that, at $z=0$, this is smaller by a factor of $\sim 2$ than the radius $r_{200,\,m}$ defined with respect to the mean matter density, as used e.g.~by \citet{Wetzel_et_al_2012}.}, and have been outside $5\, r_{200}$ in at least one preceding snapshot. This implicitly excludes galaxies which are destroyed at $z > 0$ ($\sim 40$ per cent of our original sample): our focus here is on studying the origin of the quenched population observed at the present epoch. We exclude $\sim 900$ galaxies  ($\sim 25$ per cent) affected by pre-processing --- i.e.~those which are identified as a satellite galaxy of another group or cluster with $M_\text{bound} \geq 10^{13} \msun$ --- because their simultaneous influence by two environments makes it more difficult to understand their transformation. However, we have verified that e.g.~the fraction of stripped interstellar matter (ISM) remains nearly unchanged when pre-processed galaxies are included (Section \ref{sec:discussion}; see also \citealt{Taranu_et_al_2014}). Note that our choice of boundary between the group/cluster environment and the field accounts for direct environmental influence beyond $r_{200}$, which, as we have shown in \citet{Bahe_et_al_2013}, can persist out to $\sim$ 5 times this radius. The final satellite sample includes 1884 galaxies in the stellar mass range we are interested in, i.e.~$\log_{10} (\mstar / \msun) = [9.0 - 11.0]$ at z = 0.

We also form a complementary sample of \emph{field} galaxies, which are centrals never found within $5 \,r_{200}$ from any group or cluster, in any snapshot; they will frequently be of use as a control sample. \textsc{gimic} includes a somewhat larger number of these ($\sim$ 3200 at $z = 0$). Note that our selection explicitly excludes contamination from `ejected' group/cluster satellites (see e.g.~\citealt{Wetzel_et_al_2014}), allowing us to cleanly compare galaxies affected by the group/cluster environment with those that have not.


\section{Star formation quenching in GIMIC}
\label{sec:trends}

\subsection{Low-redshift sSFR distributions}

\begin{figure*}
  \centering
	\includegraphics[width=2.1\columnwidth]{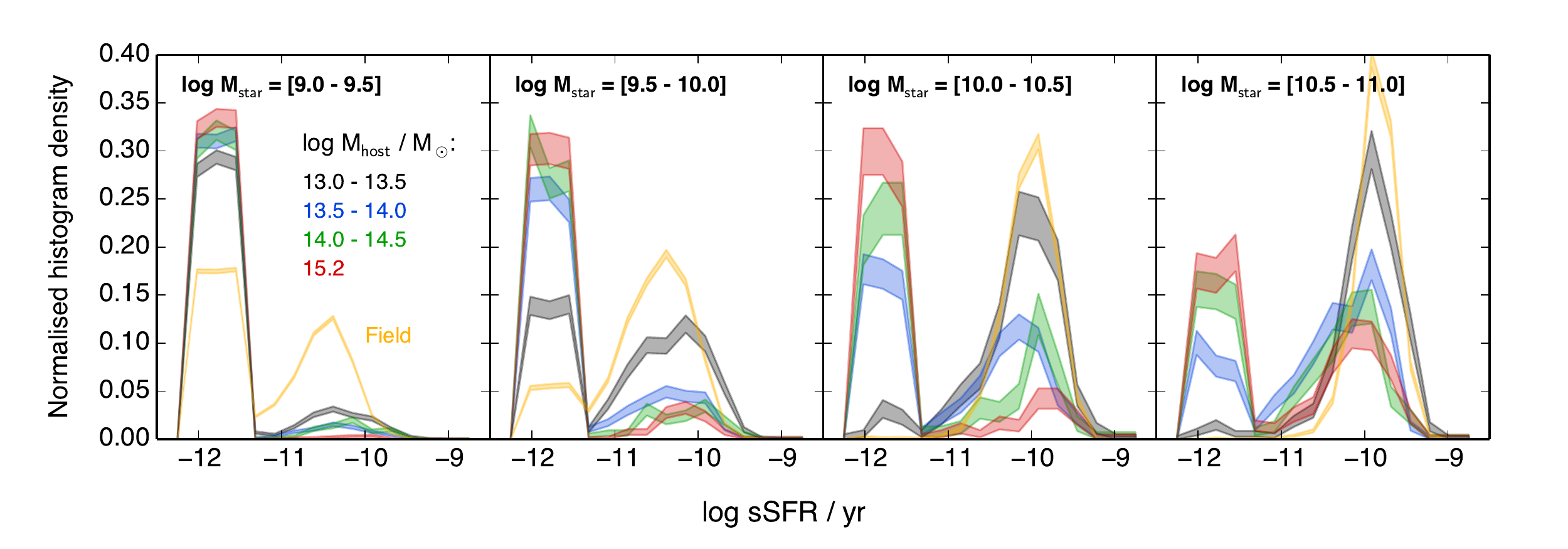}
	\caption{Distribution of specific star formation rates (sSFR) at $z \leq 0.2$ in different environments. Individual panels contain galaxies of similar stellar mass (see top), different environments are represented by different colours. Band widths indicate the statistical $1\sigma$ uncertainty on the normalised histogram density. For improved clarity, galaxies with sSFR $< 10^{-11.5}$ yr$^{-1}$ are assigned (random) sSFR values between $10^{-12.25}$ and $10^{-11.5}$ yr$^{-1}$ as described in the text. In qualitative agreement with observational data, the distributions appear bimodal, the fraction of galaxies with sSFR $> 10^{-11}$ yr$^{-1}$ decreases with increasing host mass, but the location of the `active peak' remains largely unchanged. The simulations also show an increase in the star-forming fraction with increasing stellar mass that is in contrast to observations, plausibly due to lack of AGN feedback in \gimic{}.}
\label{fig:quench.ssfrspec}
\end{figure*}

We begin our analysis by showing, in Fig.~\ref{fig:quench.ssfrspec}, the distribution of specific star formation rates (sSFR $\equiv$ SFR / $M_\text{star}$) for galaxy observations at redshift $z \leq 0.2$ in different environments. The four individual panels contain galaxies of varying stellar mass, increasing from left to right as indicated near the top of each panel. Different environments are encoded by differently coloured lines: Field galaxies are shown in yellow, while satellites at $r \leq 2\,r_{200}$ from the host centre are represented by bands in black, blue, green and red, according to the mass of their host halo (see key in the left-most panel). This radial cut was chosen to approximately match the selection of \citet{Wetzel_et_al_2012}, but other choices (such as 1 or 3 $r_{200}$) lead to qualitatively similar results. Many galaxies in our simulation have a star formation rate of zero; to show these in our plot without creating an artificially large `passive peak' at one arbitrary value, we randomly assign an sSFR between $10^{-12.25}$ and $10^{-11.5}$ yr$^{-1}$ to each galaxy with sSFR $< 10^{-11.5}$ yr$^{-1}$.

The sSFR distribution in Fig.~\ref{fig:quench.ssfrspec} is clearly bimodal, with an `active' (blue) peak centred around 10$^{-10}$ yr$^{-1}$ and a second population of `passive' (red) galaxies at much lower values. This is in good qualitative agreement with observations (e.g.~\citealt{Balogh_et_al_2004a,Kauffmann_et_al_2004,Wetzel_et_al_2012}), which also find a bimodal distribution with a break or `green valley' at sSFR $\approx 10^{-11}$ yr$^{-1}$ (but see \citealt{Schawinski_et_al_2014}). We note, however, that the `bimodality' is in both cases at least partly artificial, because neither our simulations nor observations can reliably determine low star formation rates (e.g.~\citealt{Brinchmann_et_al_2004}). In the following, we define `active' or `star forming' galaxies as those with sSFR $\geq 10^{-11}$ yr$^{-1}$, and refer to others as `passive'\footnote{The SFR resolution limit of the \gimic{} simulations is $\sim 10^{-2.5}\, \msun$ yr$^{-1}$, which is below this threshold even for the lowest mass galaxies in our sample.}. 

Observations also show a strongly declining fraction of star-forming galaxies with increasing stellar mass (e.g.~\citealt{Peng_et_al_2010, Wetzel_et_al_2012}), which is clearly in conflict with our simulations: these instead predict an \emph{increasing} star-forming fraction. The lack of this `mass-quenching' effect \citep{Peng_et_al_2010} is likely related to the absence of AGN feedback in the \gimic{} simulations. While this is regrettable, \citet{McCarthy_et_al_2012b} show that the star formation rates \emph{of star-forming galaxies} in \gimic{} agree well with observations up to $\mstar \approx 10^{11} \msun$. This suggests that galaxies not affected by mass-quenching --- which do exist in the real Universe --- are modelled relatively well, there are just too many of them. Additionally, there is strong observational evidence that the influences of galaxy mass and environment are separable \citep{Peng_et_al_2010}, so that the simulations can nevertheless give valuable insight into the action of the latter despite its failure in reproducing the former. 

At fixed stellar mass (within individual panels in Fig.~\ref{fig:quench.ssfrspec}), the simulations do show the expected trend towards a lower fraction of star forming galaxies with increasing \emph{host} mass (e.g.~\citealt{Wetzel_et_al_2012}), and a corresponding increase in the fraction of passive galaxies. The location of the `active peak' in our simulations is nearly unchanged between environments, as seen in observations \citep{Wetzel_et_al_2012}. In contrast to what is observed, massive simulated galaxies ($M_\text{star} \gtrsim 10^{10}\, \msun$) do show a slight excess in the `green valley' at sSFR $\approx 10^{-11}$ yr$^{-1}$; we discuss this further below.

\subsection{When is star formation quenched?}
\begin{figure*}
  \centering
	\includegraphics[width=2.1\columnwidth]{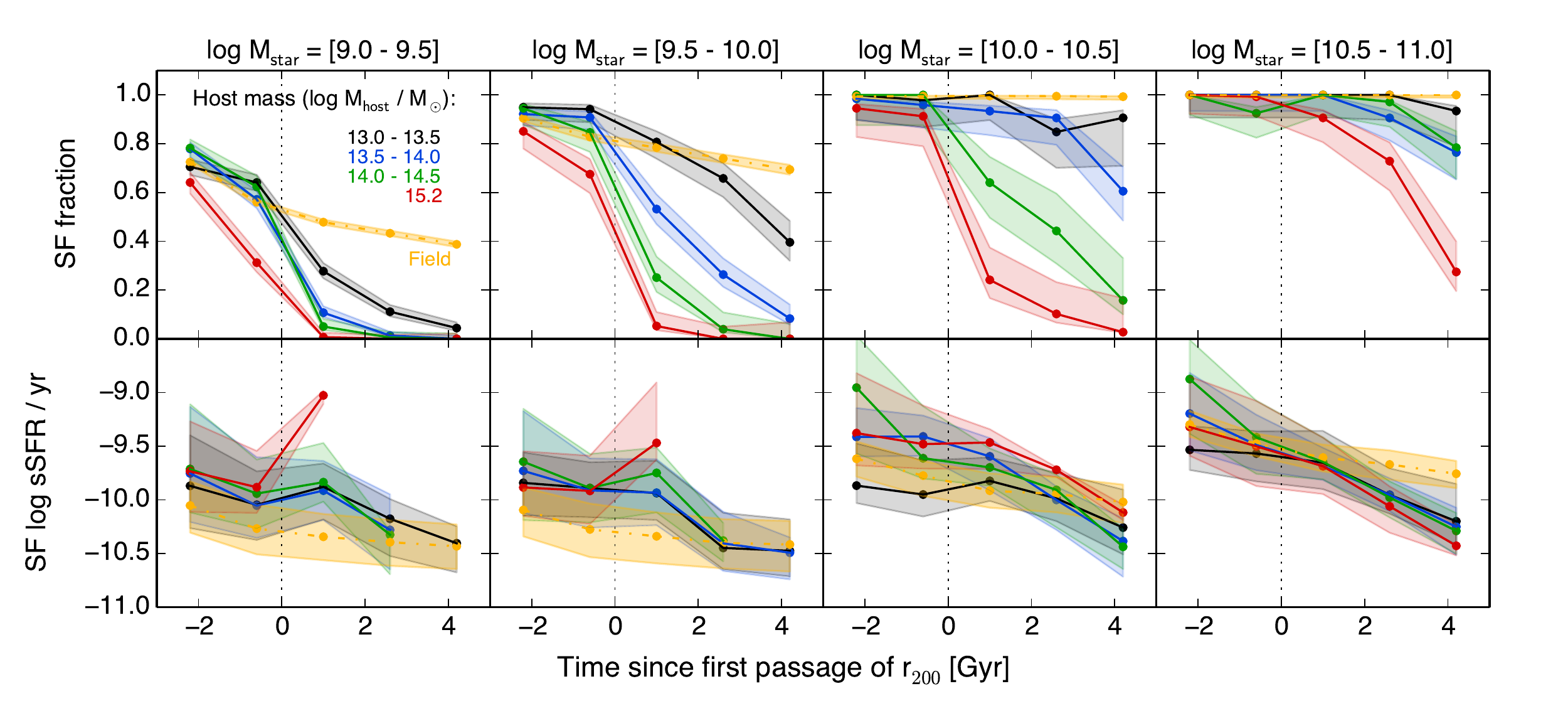}
	\caption{Evolution of star formation around the first crossing of $r_{200}$ (`accretion'). The \textbf{top} row shows the fraction of galaxies with sSFR $> 10^{-11}$ yr$^{-1}$, and the \textbf{bottom} row the median sSFR of these `star forming' galaxies. Individual columns contain galaxies of similar stellar mass as indicated on top, different colours represent differently massive hosts with field galaxies shown in yellow for comparison (see text for more details). Shaded bands represent binomial $1\sigma$ uncertainties in the top row \citep{Cameron_2011} and the 25$^\text{th}$/75$^\text{th}$ percentiles in the bottom. While the \emph{fraction} of star forming galaxies decreases strongly around accretion, the sSFR of the star forming galaxy population does not differ substantially from the field, with the exception of massive galaxies long after accretion.}
\label{fig:quench.sfrdecline}
\end{figure*}

As a first step to understanding the origin of this environmental influence, we show in Fig.~\ref{fig:quench.sfrdecline} the \emph{evolution} of star formation in infalling satellite galaxies around the first crossing of $r_{200}$; this event will henceforth be referred to as `accretion'. Galaxies are again split by their stellar mass and the mass of their host halo. Both these quantities are measured at $z = 0$ to ensure that galaxies retain the same bin designation throughout their evolution, and to directly connect the results to the observed low-redshift population. This convention is followed for the remainder of our paper except where specifically noted otherwise. 

The top row of Fig.~\ref{fig:quench.sfrdecline} shows the fraction of all satellite galaxies which are star forming, plotted against time $\Delta t$ since accretion. Compared to plotting trends with radius (as we have done in \citealp{Bahe_et_al_2013}), this has the advantage of explicitly avoiding a contamination from overshot (or `backsplash') galaxies that are no longer infalling for the first time, as well as directly revealing the relevant timescales for the transformation of the galaxy populations. However, this choice introduces an implicit redshift trend --- galaxy observations with larger $\Delta t$ are on average taken at a later point in cosmic history. It is well-known that the global sSFR is declining with time since $z \approx 2$ (e.g.~\citealp{Madau_et_al_1998,Hopkins_Beacom_2006,van_de_Voort_et_al_2011}), so one should expect to see a decline with $\Delta t$ even in the absence of \emph{any} environmental influence. To account for this, the corresponding evolution of field galaxies is included in Fig.~\ref{fig:quench.sfrdecline} as yellow dash-dot lines. For this purpose, we compute in each bin of $\Delta t$ and $\mstar$ the median redshift $z_\text{bin}$ of all group/cluster galaxies and then select all `field' observations (see Section \ref{sec:tracing}) at redshift $z_\text{obs}$ satisfying $|z_\text{obs} - z_\text{bin}| \leq 0.1$ that fall in the respective range of $\mstar$. 

The star forming fraction of satellites shows a strong decline around the time of accretion (vertical dotted line). This effect is clearly larger than the general decline of simulated field galaxies classified as star forming --- the group/cluster sample agrees well with the field $\sim 2$ Gyr prior to accretion, but then its star forming fraction shows a much stronger subsequent drop. Quenching is strongest, and occurs earliest, for low-mass galaxies (left column) and those in the massive cluster (red line), in many cases well outside $r_{200}$. Around 2 Gyr after accretion, the star forming fraction is significantly below the field in almost all combinations of galaxy and host mass, apart from the extreme case of massive galaxies in groups ($\mstar \geq 10^{10}\, \msun$ and $M_\text{host} \leq 10^{14}\, \msun$). 

The sSFR of star forming galaxies (bottom row of Fig.~\ref{fig:quench.sfrdecline}; bands here enclose 50 per cent of galaxies in each bin) shows a different behaviour: most are significantly above our adopted threshold of $10^{-11}$ yr$^{-1}$, and in virtually all bins of galaxy and host mass the sSFR decline is largely the same as in the field. This explains the constancy of the blue peak location found in Fig.~\ref{fig:quench.ssfrspec}, and suggests that most galaxies are quenched rapidly in our simulations (we will test this explicitly in Section \ref{sec:quenchingspeed}). Massive satellites, with $\mstar \gtrsim 10^{10.5} \msun$, that are still forming stars a few Gyr after accretion are an exception, and do show a somewhat lower sSFR than in the field (by about 0.5 dex at $\Delta t = 4$ Gyr). This leads to the excess of satellite galaxies on the `blue slope of the green valley' in the right-most panel of Fig.~\ref{fig:quench.ssfrspec}. It is important to keep in mind, however, that the number of star forming satellites in this mass range is artificially enhanced in our simulations: the overall effect on real galaxies is therefore necessarily smaller.

\begin{figure}
  \centering
    \includegraphics[width=\columnwidth]{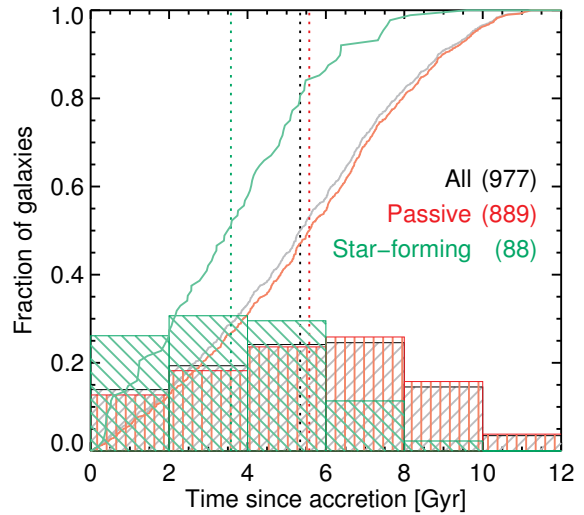}
    \caption{Distribution of galaxy accretion times (first crossing of $r_{200}$). The hatched histograms show the differential distribution, while the solid and dotted lines give the cumulative distribution and median, respectively. Colours indicate star-formation activity at $z = 0$. Galaxies were accreted over a large time interval with median $\sim 5$ Gyr ago. The (small) fraction of galaxies still forming stars at the present day were preferentially accreted later, by typically $\sim 2$ Gyr. Note that galaxies that have not yet crossed $r_{200}$ at $z = 0$ are not shown here.}
  \label{fig:quench.acctimes}
  \end{figure}

Fig.~\ref{fig:quench.sfrdecline} also appears to show a (temporary) \emph{increase} in sSFR shortly after accretion, in particular for lower-mass galaxies in the massive cluster (red lines). This is actually an artefact introduced by implicit redshift dependence: As Fig.~\ref{fig:quench.acctimes} shows, galaxies which are satellites at redshift $z = 0$ were accreted over a broad time interval of $\sim 10$ Gyr, with a peak $\sim 6$ Gyr ago (see also \citealt{Wetzel_et_al_2013, Taranu_et_al_2014}). Galaxies accreted earlier were able to continue forming stars for a longer time after accretion, because the host halo was less massive then. The sSFR was also generally higher at higher $z$, which causes the apparent increase when stacking the evolution of galaxies accreted across cosmic history. Despite this, \emph{currently} star-forming satellite galaxies (coloured green in Fig.~\ref{fig:quench.acctimes}) were preferentially accreted later --- by typically almost 2 Gyr --- as expected in a scenario where the likelihood of star formation being quenched increases with time since accretion \citep{Wetzel_et_al_2013}. 

\section{The fate of star forming gas in quenched satellite galaxies}
\label{sec:reason}
To find the mechanism(s) responsible for environmental star formation quenching, we now focus on those galaxies which are passive and within $5\,r_{200}$ from the host centre at $z = 0$: this is the case for the majority of our group/cluster galaxies (1402 out of 1884, including galaxies that have not yet crossed $r_{200}$ at $z=0$ and are therefore not included in Figs.~\ref{fig:quench.sfrdecline} and \ref{fig:quench.acctimes}). In the simulations, stars are formed from dense gas with $n_\text{H} \geq 0.1 \text{cm}^{-3}$ (see \citealt{Schaye_DallaVecchia_2008}), which we can identify as the galaxy's interstellar matter (ISM). Galaxies with little to no star formation must therefore be lacking this high-density ISM gas, and determining the reason for their passive nature means finding out why this is the case.

By definition, all galaxies contain stars which must have formed at some point in the past. It is therefore unlikely that galaxies which are passive at $z = 0$ have been devoid of ISM throughout cosmic history: rather, they must have lost it at some point in their life. There are two broad scenarios for how this could have happened: Either, the gas was used up internally --- by star formation and/or expulsion through galactic winds --- and not replenished, or it was stripped by external forces. To distinguish between these different scenarios, we will exploit the Lagrangian nature of the SPH formalism adopted for the \gimic{} simulations: the particles making up each galaxy's ISM in one `initial' snapshot at redshift $z_i$ can be identified again in a subsequent snapshot at redshift $z_s < z_i$. We can then assign each particle to one of four categories based on its status at $z_s$: (i) still part of the galaxy's ISM; (ii) turned into stars; (iii) still bound to the galaxy, but no longer sufficiently dense to qualify as ISM; or (iv) having left the galaxy's subhalo altogether. The fraction of ISM in each category can then be used as a `fingerprint' to infer the physical mechanisms that have led to its depletion.

In choosing these two snapshots $z_i$ and $z_s$, we need to consider two constraints. Firstly, quenching is no instantaneous process and may, in some cases, be dragged out over a period of several Gyr (c.f.~our discussion of Fig.~\ref{fig:quench.sfrdecline} above). It is plausible that some ISM gas may be replenished during this time, so that we cannot just analyse the fate of the ISM in any \emph{one} snapshot to find the mechanism responsible for the overall ISM depletion. Therefore, we find for each galaxy a `last normal' snapshot (our procedure for this will be discussed shortly), and then take all subsequent observations of this galaxy at $z > 0$ as initial snapshots for our fate determination. This ensures that we capture the fate of \emph{all} particles that make up the galaxy's ISM at any point during the quenching process. Secondly, a gas particle may undergo more than one transformation between the initial redshift $z_i$ and $z = 0$; for example, it may move to the extended halo, remain there for some time, and then be removed from the galaxy. This is clearly not the same as gas being stripped immediately from the star-forming disk. To distinguish as best as possible between such cases, we therefore choose our subsequent snapshot to be the one immediately following each of our `initial' snapshots, typically $\sim$ 250 Myr later.


\subsection{Determining the onset of star formation quenching}
\label{sec:onsetdetermination}
Our analysis method outlined above requires the determination of a point of quenching onset, which marks the beginning of environmental influence on a galaxy's (specific) star formation rate. In the past, many studies have chosen this starting point as the time when the galaxy is accreted onto a larger halo, defined e.g.~as having crossed some characteristic radius such as $r_{200}$ (e.g.~\citealt{McGee_et_al_2009,Wetzel_et_al_2013}). However, we have shown in \citet{Bahe_et_al_2013} that the group/cluster environment can affect galaxies significantly earlier than this (see also Fig.~\ref{fig:quench.sfrdecline}), so that we risk missing at least part of the transformation process if we were to only consider the evolution of galaxies after accretion. Instead of this traditional `halo-based' definition, we therefore adopt a `galaxy-based' approach, and define the onset of quenching as the point when a galaxy's sSFR begins to fall below the value exceeded by 84 per cent of  comparable field galaxies (i.e.~deviates by more than $1\sigma$), \emph{independent of its location} in the group or cluster. In this way, we are guaranteed to capture the entire ISM depletion process, irrespective of its start- or endpoint. 

\begin{figure}
  \centering
    \includegraphics[width=\columnwidth]{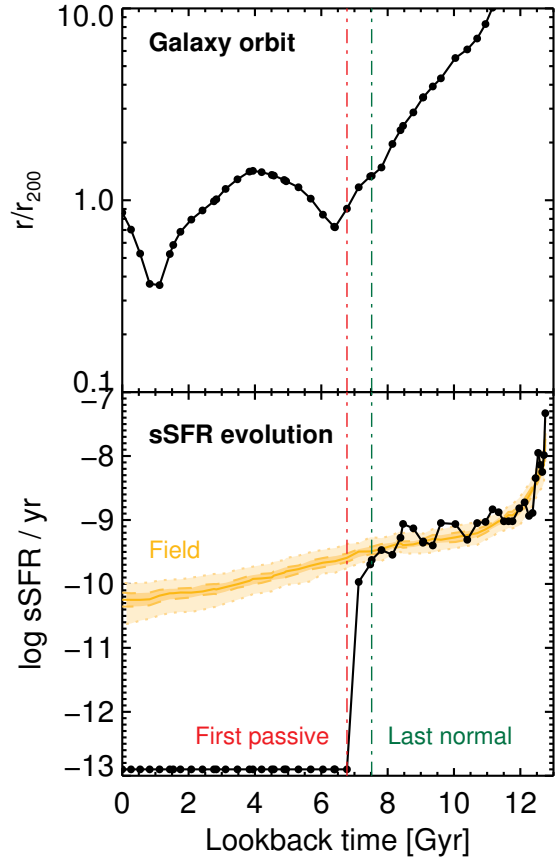}
    \caption[Example galaxy evolution]{Evolution of a galaxy with $\mstar \approx 10^{10}\, \msun$ falling into a massive cluster with $M \approx 10^{15.2}\, \msun$. The \textbf{top} panel shows the galaxy's cluster-centric radius $r / r_{200} (z)$, the \textbf{bottom} panel its sSFR (black lines, circles show individual observations of this galaxy). The yellow line (median) and shaded bands (enclosing 50 and 68.2 per cent of comparison galaxies, respectively) in the bottom panel represent the sSFR distribution of field galaxies with similar $\mstar (z)$. Environmental star formation quenching becomes apparent when the sSFR of the galaxy drops below the field (green vertical line). The red line marks the galaxy's transition from star-forming to passive $\sim 1$ Gyr later, shortly before its first pericentric passage.}
  \label{fig:quench.galtrack}
  \end{figure}

Fig.~\ref{fig:quench.galtrack} illustrates this procedure for one representative galaxy with $\mstar \approx 10^{10} \,\msun$ in the massive cluster ($M_\text{bound} = 10^{15.2}\, \msun$). The top panel shows the evolution of its cluster-centric radius $r/r_{200}\, (z)$ over cosmic time (from right to left), and the bottom panel the corresponding evolution in sSFR (solid black lines, filled circles indicate individual snapshots). The latter is overall declining since high redshift, interrupted by several short bursts where it is temporarily increased. Shortly before the first pericentric passage $\sim 7$ Gyr before today, star formation stops completely (red vertical line), plotted here at an arbitrary level of $10^{-13} \text{ yr}^{-1}$. 

In part, this decline likely reflects the general decrease of the cosmic sSFR with time since $z \approx 2$ (e.g.~\citealt{Hopkins_2004}). To isolate the \emph{environmental} influence, we find for each observation of this cluster galaxy a sample of similar field galaxy observations, defined as those deviating by less than 0.05 in observation redshift\footnote{We allow for a small redshift discrepancy because the snapshots of the individual \gimic{} simulations are not all output at exactly the same redshift.}, and 0.1 in log ($\mstar /\msun$). The sSFR distribution of this matched field sample is shown in yellow in Fig.~\ref{fig:quench.galtrack}, the light shaded band enclosing the $1\sigma$ scatter region (68.2 per cent of the sample). We can then define the `last normal snapshot', and thus the onset of environmental star formation quenching, as the point at which the satellite's sSFR falls below this $1 \sigma$ field distribution, here at a lookback time of $\sim 7.5$ Gyr (green vertical line). Note that the galaxy has not yet come within $r_{200}$ at this point. Although this particular choice of threshold is to some extent arbitrary, we have verified that other plausible definitions --- such as the point when the sSFR drops below the median or 37.5$^\text{th}$ percentile of the field distribution --- has no significant impact on our results. 

By construction, the onset of quenching determined in this way could in principle be entirely unrelated to an individual galaxy's proximity to a group or cluster. However, we have verified that the vast majority of our galaxies have a `last normal' snapshot within $\sim 3\,r_{200}$, i.e. at radii where environmental influences are expected to be significant (\citealt{Bahe_et_al_2013}; see also Fig.~\ref{fig:quench.rpprofile} below).


\subsection{How fast are galaxies quenched?}
\label{sec:quenchingspeed}
\begin{figure}
  \centering
	\includegraphics[width=\columnwidth]{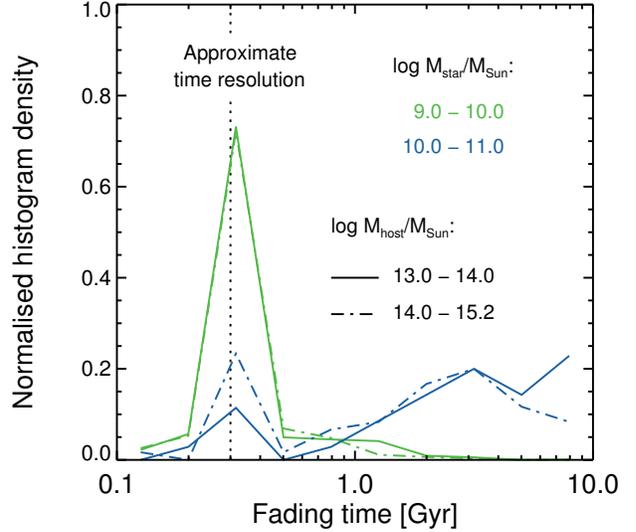}
	\caption{Distribution of fading times for galaxies of different stellar masses, defined as the interval between the last observation with sSFR consistent with the field, and the first one with sSFR $< 10^{-11}$ yr$^{-1}$. Quenching is rapid for low-mass simulated galaxies (green), but often much more protracted for more massive ones (blue). The host mass, on the other hand, appears to have relatively little influence on this quantity.}
\label{fig:quench.dtd}
\end{figure}

Having identified the time of quenching onset, we can directly determine how long each individual galaxy took from this point until it became passive (sSFR $< 10^{-11}$ yr$^{-1}$); we refer to the length of this interval as the `fading time' (as in \citealt{Wetzel_et_al_2013}). In Fig.~\ref{fig:quench.dtd} we show the distribution of this quantity for galaxies below and above $\mstar = 10^{10}\,\msun$ (green/blue lines), which are satellites in groups ($ \log \,M_\text{host}/\msun = [13.0, 14.0]$; solid lines) and clusters ($ \log M_\text{host}/\msun = [14.0, 15.2]$; dash-dot lines). The influence of galaxy mass is clearly greater: the lower mass galaxies are almost all quenched within $\leq 500$ Myr, irrespective of the mass of their host halo. For this mass range, our simulations confirm the conclusion of \citet{Wetzel_et_al_2012} that the constancy of the blue peak sSFR is due to rapid environmental quenching.

Many galaxies with $\mstar > 10^{10}\, \msun$, on the other hand, show a more gentle decline of their sSFR in the simulations, and only become passive after many Gyr of below-field-level star formation activity. This slow decline, which agrees better with the observational results of e.g.~\citet{von_der_Linden_et_al_2010} than \citet{Wetzel_et_al_2013}, is slightly more common in groups (solid lines) than clusters and suggests that the harsher environment of clusters is somewhat more conducive to rapid quenching. A slow decline explains why star-forming galaxies with $\mstar > 10^{10} \msun$ show median sSFR values slightly below the field several Gyr after accretion (Fig.~\ref{fig:quench.sfrdecline}), and the corresponding excess of galaxies with sSFR $\approx 10^{-11}$ at $z \approx 0$ seen in Fig.~\ref{fig:quench.ssfrspec}: the longer the fading time, the higher the chance of finding a galaxy in this stage of its life. We caution, however, that many of these galaxies would in reality have been quenched already as centrals prior to accretion, and it seems at least plausible that whichever mechanism is responsible for this `mass-quenching' may also shorten the fading time for those galaxies that are still forming stars at accretion. 


\subsection{Importance of different ISM loss routes}
\label{sec:ismanalysis}
We now compute the mass $m_{i\!j}$ of ISM lost between each two snapshot pairs $i$ since the `last normal' snapshot in different ways $j$. Summing over $i$ gives the total mass of ISM the galaxy has lost in each way during quenching, $M_j = \sum_i m_{i\!j}$. The fraction of ISM loss that can be attributed to each of these routes is then simply $M_j / \sum_j M_j$.

\begin{figure*}
\centering
  \includegraphics[width=2.1\columnwidth]{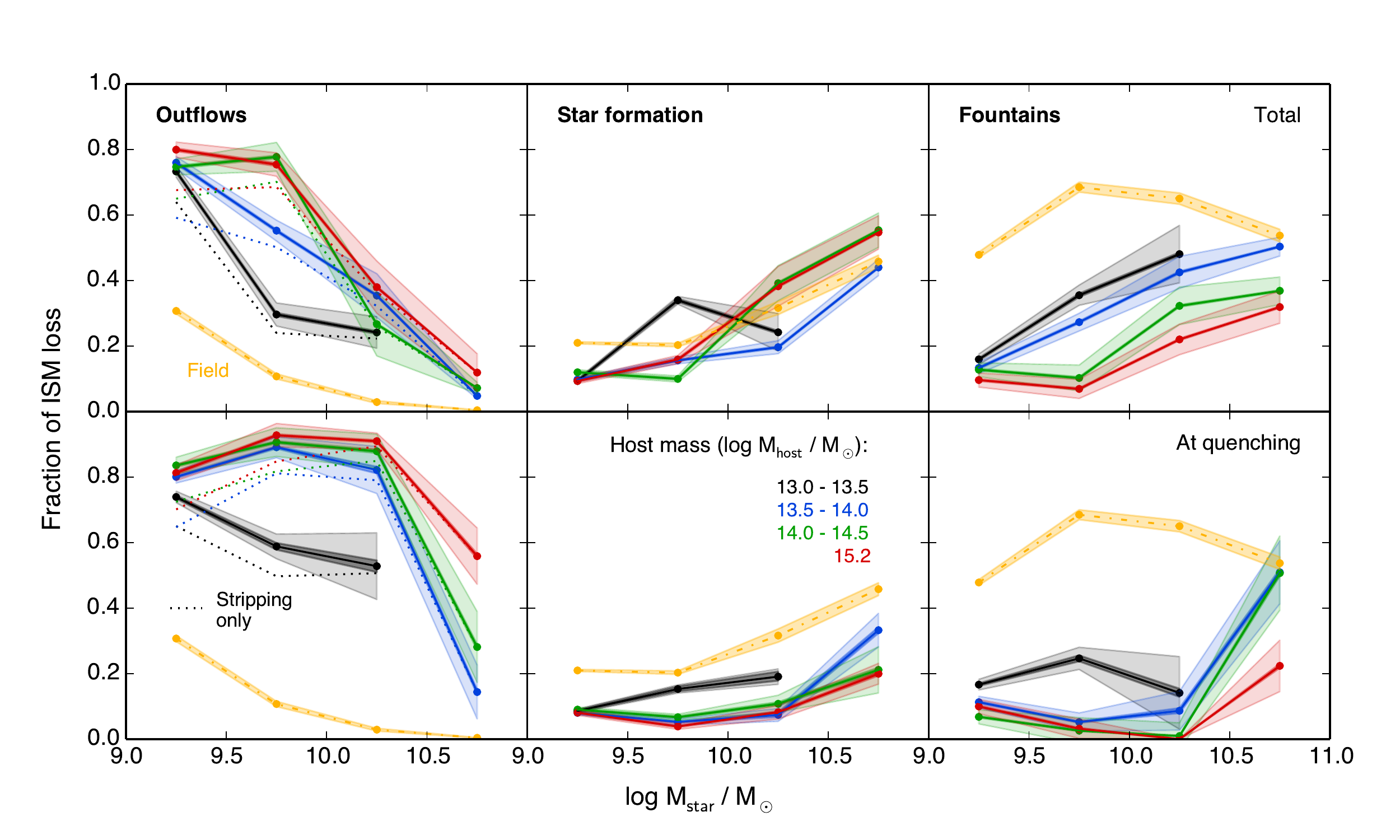}
  \caption{Fate of ISM gas in satellite galaxies in the \gimic{} simulations which are passive at $z = 0$. Shown is the average fraction of loss due to `outflows', star formation, and galactic fountains. The top line analyses the integrated loss over the whole quenching process, whereas the bottom row focuses on the loss immediately prior to quenching. Different colours represent different host masses (yellow: field), as indicated. Star formation itself only accounts for a moderate fraction of ISM removal, and outflows from the galaxy are important in particular just prior to quenching (bottom line). The dotted lines show the fraction of ISM loss due to satellite-specific mechanisms (stripping), which clearly accounts for the majority of overall outflows (solid lines).}
    \label{fig:quench.coldfate}
\end{figure*}

Fig.~\ref{fig:quench.coldfate} shows the result of this analysis. Its three columns give, from left to right, the mass fraction of ISM removed by `outflows'\footnote{We use the term `outflow' for all gas leaving the galaxy, irrespective of how it was motivated to do so (see next section).}, star formation, and `fountains' \citep{Shapiro_Field_1976}, by which we refer to gas which remains bound to the galaxy halo, but is no longer sufficiently dense to form stars. In the top row, we analyse the ISM loss averaged over the entire quenching process, while the bottom row focuses only on the snapshot interval immediately before a galaxy becomes passive. As in Fig.~\ref{fig:quench.sfrdecline}, we separate galaxies by stellar mass (here represented by different bins on the x-axis) and host mass (differently coloured lines), both measured at $z = 0$. For comparison, the corresponding fractions in the field are shown in yellow\footnote{We take the field value as the median of the comparison sample, in the top row at the snapshot halfway between `last normal' and `first passive'.}. The light shaded bands indicate the $1\sigma$ uncertainty due to galaxy-to-galaxy scatter, while the (mostly thin) dark shaded bands represent the corresponding counting uncertainty from particle-to-particle scatter. 

The first clear result from Fig.~\ref{fig:quench.coldfate} is that, integrated over the entire quenching process, no one route is dominating the removal of ISM across the whole range of galaxy and host masses that we consider. Likewise, none is completely negligible: star formation, for example, accounts for $\sim 50$ per cent of ISM loss in the highest mass galaxies, dropping to $\sim 10$ per cent at the low-mass end. Nevertheless, we see a substantial importance of `outflows' (left panel): this is particularly evident in the case of low-mass galaxies with $M_\text{star} \lesssim 10^{10}\, \msun$, where they account for $> 50$ per cent of ISM loss outside poor groups (black). In more massive galaxies, their significance is reduced, which is mirrored by an increased fraction of ISM lost through fountains and star formation. 

\subsubsection{Satellite-specific outflows}
\label{sec:envexc}
It is important to keep in mind that `outflows' as defined above can be driven both externally, for example by ram pressure stripping, and internally through processes such as supernova feedback. A simple method to find the amount of ISM removed by satellite-specific processes is to compute the \emph{`environmental excess'} of outflows in satellites compared to the field: assuming that internal outflow mechanisms lead to the same specific rate of ISM outflow $\mu \equiv \dot{M}/M$ in the field and in satellites, this excess is $x_E = (\mu_\text{sat}-\mu_\text{field})/\mu_\text{sat}$. The fraction of ISM loss in satellites caused by external influences is then $x_E \times f_\text{out}$, where $f_\text{out} = \dot{M}_\text{out}/\dot{M}_\text{ISM loss}$ is the fraction of (total) ISM loss due to outflows (solid lines in Fig.~\ref{fig:quench.coldfate}):
\begin{equation}
f_\text{env} = \frac{\mu_\text{sat} - \mu_\text{field}}{\mu_\text{sat}} \frac{\dot{M}_\text{out}}{\dot{M}_\text{ISM loss}}
\end{equation}
The result is shown with dotted lines in the left column of Fig.~\ref{fig:quench.coldfate}. Evidently, outflows in satellite galaxies are largely environment-driven, in agreement with their larger role in more massive hosts. For galaxies with $\mstar \approx 10^9\, \msun$, environmental outflows account for roughly two thirds of the ISM removal, making them the dominant quenching process. We will address their physical origin in detail in the next section.

\subsubsection{Variation during quenching}
To get an indication for how these trends vary \emph{during} the quenching period, we show in the bottom row of Fig.~\ref{fig:quench.coldfate} the corresponding fractions obtained by considering only the ISM lost immediately before the galaxy became passive, i.e.~following the last snapshot with sSFR $> 10^{-11}\,\,\text{yr}^{-1}$. The most noticeable difference is that, with the exception of the most massive galaxies ($\mstar > 10^{10.5}\, \msun$), stripping is here clearly the dominant mechanism, accounting for $\sim 80$ per cent of ISM removal in hosts with $M > 10^{13.5}\, \msun$. In contrast, the fraction of ISM recycled into the halo and even that turned into stars is significantly reduced at this point. Evidently, stripping is most important at the end of the quenching process, whereas star formation and fountains play a larger role in its early phase. This suggests an \emph{interrupted strangulation} scenario, where a relatively slow depletion of the ISM through internal consumption removes a substantial part of the ISM (particularly in galaxies with $\mstar \gtrsim 10^{10}\, \msun$), but is not fast enough to complete this process before external forces become sufficiently strong for direct stripping to take over. 


\section{The mechanisms causing satellite-specific outflows}
As we have shown above, the majority of ISM outflows from galaxies being quenched in groups and clusters are driven by mechanisms operating only on satellites. We now investigate their physical origin, beginning with the nature of external forces acting on the ISM during quenching.

\subsection{Negligible influence of tidal stripping}
\begin{figure}
  \centering
    \includegraphics[width=\columnwidth]{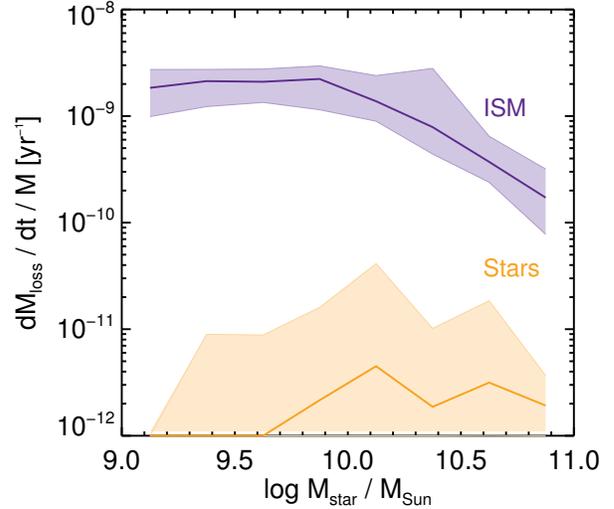}
    \caption{The rate at which ISM and stellar mass in the central 20 kpc of satellite galaxies is removed during quenching of star formation. Shown is the median specific loss rate of galaxies as a function of $\mstar$ (bands enclose 50 per cent of galaxies). The value for each individual galaxy is the average over all snapshots since the sSFR dropped below the field, weighted between snapshots by the amount of instantaneous ISM loss. The ISM is affected orders of magnitude more strongly than stars, implying that tidal forces --- which would remove both alike --- are not the dominant mechanism for its depletion.}
    \label{fig:quench.stellarstripping}
  \end{figure}

Satellite galaxies orbiting in groups and clusters are subject to both hydrodynamical effects, in particular ram pressure, and tidal forces. While both have a tendency to remove matter from the satellite, one key difference between them is that ram pressure affects only the gaseous component, while tidal forces strip matter of any kind, including stars. In Fig.~\ref{fig:quench.stellarstripping}, we therefore compare the specific loss rate $\dot{M}_\text{loss}/M$ of ISM gas (purple) to that of stars (within 20 kpc from the galaxy centre to crudely exclude the stellar halo; yellow). $\dot{M}_\text{loss}$ is the rate at which gravitationally bound mass leaves the galaxy. The loss of ISM is stronger by a factor of $\sim 100$ at the high-$\mstar$ end of our galaxy sample, and even more at lower masses. This strongly suggests that \emph{tidal stripping has negligible influence on the removal of ISM in our simulated galaxies}. 

Using a simple analytic argument, \citet{McCarthy_et_al_2008b} reached a similar conclusion, finding that tidal stripping of gas should only be important when the mass of the satellite is comparable to that of the host halo (ratio larger than $\sim$ 1:8). We emphasise that this does \emph{not} mean that tidal stripping has no effect at all on group and cluster galaxies: indeed, in order to explain the relatively large fraction of stars residing in the diffuse intracluster light component at $z = 0$ in both observations (e.g.~\citealt{Gonzalez_et_al_2005,Gonzalez_et_al_2007,McGee_Balogh_2010,Budzynski_et_al_2014,DSouza_et_al_2014}) and simulations (e.g.~\citealt{Puchwein_et_al_2010,Cui_et_al_2014}), tidal stripping must eventually become important, but generally well after the cold ISM has been removed by hydrodynamic processes\footnote{Because our sample only includes galaxies identified at $z = 0$, it is in principle possible that some ISM is removed by tidal forces as part of the complete disruption of a satellite. The analytic argument of \citet{McCarthy_et_al_2008b} makes this quite unlikely, however, and in any case it would be irrelevant in explaining the origin of quenched \emph{surviving} galaxies.}. 

\subsection{Ram pressure stripping}
The argument above suggests that ram pressure is the key driver of satellite-specific outflows, although it is so far mostly a proof by exclusion. To verify its importance, we can use the spatial information provided by the simulations, and consider \emph{in which direction} ejected gas particles are leaving the galaxy. Gas removed through ram pressure should be distributed in a tail pointing away from the galaxy's motion with respect to the ICM. Tidal stripping will lead to a double-tail structure (one leading, one trailing), while gas ejected through internal feedback should show no preferential direction with respect to the surrounding ICM. 

To apply this method, we need to determine the direction of the galaxy's motion with respect to the ICM, or equivalently the velocity of the ICM in the galaxy's rest frame. As in \citet{Bahe_et_al_2013}, we use a simple and robust method to accomplish this: in each snapshot, we find the 3000 gas particles nearest each galaxy centre which have a density $n_\text{H} \leq 0.01 \text{ cm}^{-3}$, are not currently gravitationally bound to any subhalo except the main group/cluster, and are not bound to the galaxy in any snapshot. This avoids biases due to small dense clumps of gas, orbiting satellite galaxies, and gas falling in or flowing out from the galaxy itself, with a number of particles large enough to yield statistically robust results. We then take the ICM velocity $\mathbf{v}_\text{ICM}$ as the mass-weighted average of the velocity in the galaxy rest frame $\mathbf{v}_i$ of these 3000 particles (i.e. $\mathbf{v}_\text{ICM} = \sum_i{\textbf{v}_i m_i}/\sum_i{m_i}$). For each particle which has been lost from the galaxy since the preceding snapshot, we can then determine its position angle relative to this velocity vector, $\cos \gamma = \mathbf{r}_p \cdot \mathbf{v}_\text{ICM} / (| \mathbf{r}_p||\mathbf{v}_\text{ICM}|)$, where $\mathbf{r}_p$ is the galacto-centric position of the particle.

\begin{figure}
  \centering
    \includegraphics[width=\columnwidth]{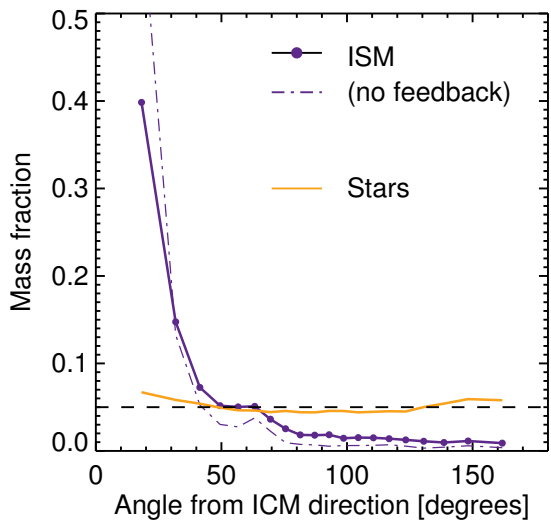}
    \caption{The distribution of position angles $\gamma$ with respect to the direction of the ICM velocity (in the galaxy frame) for particles lost from satellite galaxies. ISM particles are shown in purple (solid line: all, dash-dot line: only those not directly affected by feedback), stars in orange. There is a clear excess of ISM particles at small angular separation, as expected if these particles were stripped by ram pressure, as well as a `floor' of particles at large $\gamma$. Note that there is \emph{no} sign of an enhancement at large angles, which would be expected from tidal stripping. The distribution for stars, on the other hand clearly shows this `double-tail' feature.}
    \label{fig:quench.multang}
  \end{figure}

The resulting distribution of $\gamma$ is shown in Fig.~\ref{fig:quench.multang}; for simplicity we have stacked together all galaxies in our quenched sample. Particles are grouped into 20 bins of equal solid angle in the first snapshot after having left the galaxy, so that in a completely random distribution, each bin would contain 5 per cent of them (dashed horizontal line). However, this is clearly not what we actually find: There is a very strong excess of ISM at small angles ($\gamma \lesssim 50^{ \circ}$), as expected for ram pressure stripping (solid purple line). At larger angles, there are fewer particles than in a random distribution, although it never drops completely to zero. In agreement with expectations from Fig.~\ref{fig:quench.coldfate}, this shows that \emph{some} ISM outflow in satellites is purely the result of supernova feedback, rather than ram pressure stripping. For comparison, we also show the distribution for only those ISM particles which have not been directly affected by SN feedback (dash-dot line, see below). Not surprisingly, this shows a stronger peak at small angles, and drops to virtually zero for $\gamma \gtrsim 90^\mathrm{o}$. 

There is no sign of an increased fraction of ISM particles emitted at large angles --- which would be indicative of a `leading' tail produced from tidal stripping --- confirming the expectations from Fig.~\ref{fig:quench.stellarstripping} above. In contrast, the direction of \emph{star particles} lost from the galaxy (shown as orange line, note that this includes stars lost after removal of the ISM) shows a weak but clear double-peak structure, with a trailing and leading tail separated by a minimum around 90 degrees. Together, these features strongly support the interpretation that ram pressure is the key driver of satellite-specific outflows.

\subsubsection{Comparison to `satellite-specific' outflows}
To test the importance of ram pressure stripping in a quantitative way, we can compute the fraction of ISM lost into a (trailing) tail, and compare this to the `environmental excess' determined above. Motivated by the assumption that particles stripped by ram pressure will always be emitted into a trailing (i.e.~$\gamma < 90^\circ$), we separate all ISM particles lost through outflows into those with $\gamma < 90^\circ$ (fraction $f_1$) and the complementary set with $\gamma > 90^\circ (f_2)$. A certain fraction $\alpha$ of these particles is stripped by ram pressure; by our above assumption these will all end up in $f_1$. The remaining fraction $\beta = 1 - \alpha$ is removed by feedback, and will equally likely end up in $f_1$ or $f_2$. The fraction of outflows accounted for by ram pressure stripping is then $\alpha = f_1 - f_2$. 

\begin{figure*}
\centering
  \includegraphics[width=2.1\columnwidth]{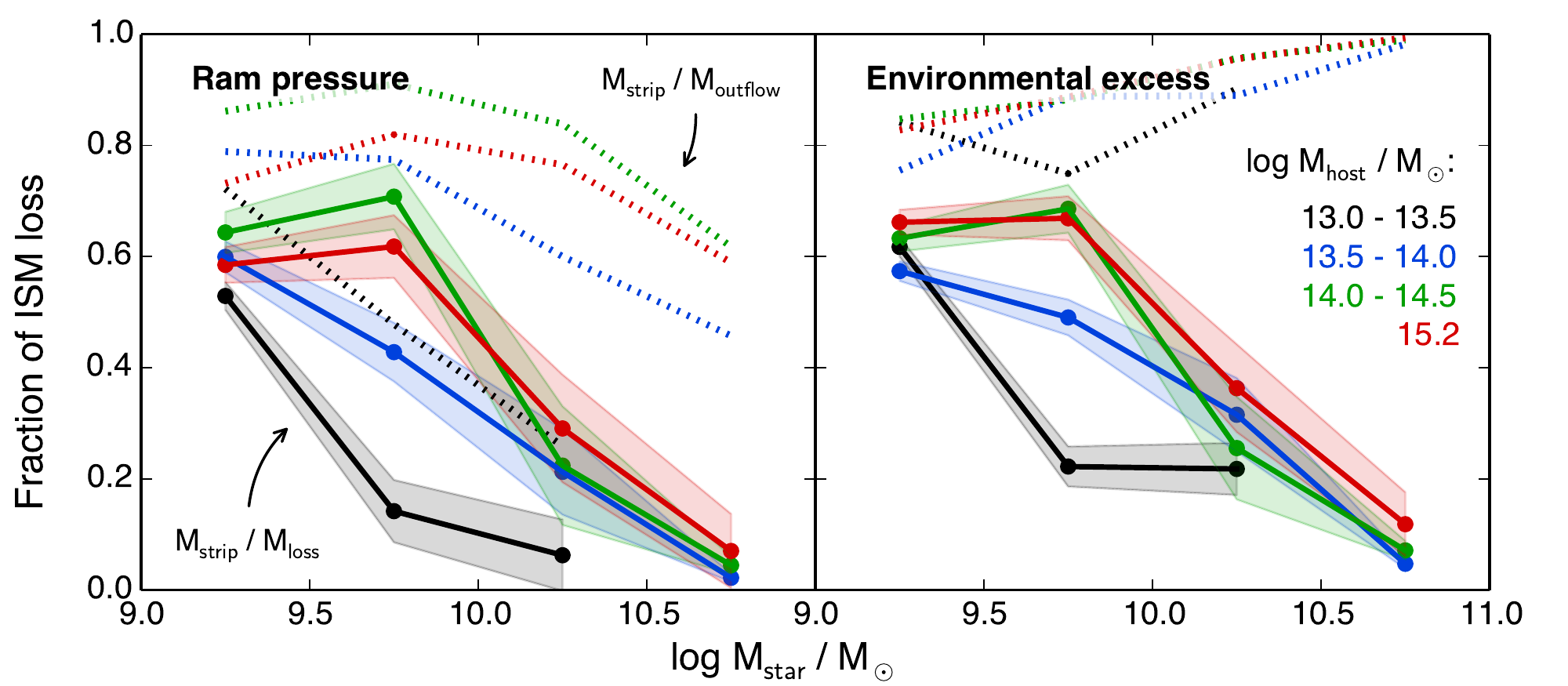}
 \caption{Comparison of the fraction of ISM lost through ram pressure stripping (left) and satellite-specific processes in general (right). Lines of different colour represent galaxies in differently massive hosts. Solid lines give the fraction out of the total ISM loss, and dotted lines out of gas lost through outflows (which is always higher by definition). Shaded bands represent the statistical $1\sigma$ uncertainties from galaxy-to-galaxy variation; for clarity these are only plotted for the first set. Both methods agree well, indicating that ram pressure is the key mechanism at work in removing the ISM from satellite galaxies which are quenched at $z = 0$.}
  \label{fig:quench.variation}
\end{figure*}

In Fig.~\ref{fig:quench.variation}, we compare the result of this \emph{`tail method'} (left panel) with our above-determined `environmental excess' (right). In both cases, solid lines represent fractions relative to the total loss of ISM from the galaxy, in the same way as in Fig.~\ref{fig:quench.coldfate}. The dotted lines show fractions relative to only the total loss through outflows, which are necessarily higher.

At low stellar masses, the two methods agree very well, both accounting for approximately 80 per cent of outflows, and 60 per cent of the total ISM loss. Towards more massive galaxies, on the other hand, the fraction of outflows that are satellite-specific (right panel, dotted line) increases with stellar mass, whereas the fraction removed by ram pressure into a tail (left panel) decreases, particularly in galaxy groups. In other words, there must be an additional satellite-specific effect causing (some) outflows of ISM gas, beyond ram pressure stripping. A possible explanation is that feedback-driven outflows are more effective in escaping from satellites than in the field, for which there could be (at least) two reasons. Firstly, the potential wells of satellite galaxies may be less deep than those of centrals with the same stellar mass, due to partial removal of their dark matter. Secondly, as we show below, satellites have already lost a substantial part of their gaseous halo by the time quenching begins, which reduces the possibility of confining outflows from the disk through hydrodynamic interactions. Both effects would increase the outflow rate in satellites without leading to a tail structure, and could therefore explain the discrepancy between the two panels of Fig.~\ref{fig:quench.variation}. It is also clear, however, that this effect is strongest at the high-mass end of our sample, where the total importance of ISM outflows is rather small. \emph{In conclusion, ram pressure stripping accounts for the vast majority of satellite-specific ISM outflows, and the majority of \emph{all} ISM loss in satellites with $\mstar < 10^{10}\, \msun$ in hosts with masses as low as $10^{13.5}\, \msun$}.

\subsection{Interplay between feedback and stripping}
We have shown above that a substantial fraction of ISM is driven out of group and cluster satellites by ram pressure during quenching. Naturally, this must lead to a reduced fraction of ISM lost otherwise. Fig.~\ref{fig:quench.coldfate} shows that this primarily affects the fraction returned to the halo through fountains, while --- at least averaged over the whole quenching period --- the fraction of ISM turned into stars is similar to the field. This suggests that at least some stripped ISM has \emph{first} been expelled into the halo through feedback, and was then stripped almost immediately from there. Compared to stripping of gas directly from the disk (as envisaged by \citealt{Gunn_Gott_1972}), this `stripping of fountains' requires lower levels of ram pressure, because the gas is less tightly bound as well as less dense. In this scenario, the removal of ISM is therefore the result of a synergy between feedback and ram pressure stripping. 

\subsubsection{Stripping from the disk and from fountains}
We test this hypothesis with the ram pressure model proposed by \citet{Gunn_Gott_1972}, by determining how much ISM could be stripped directly from the disk. Any potential shortfall between this and the total stripped fraction (as shown in Fig.~\ref{fig:quench.variation}) must then be due to gas removed by stripping of fountains. In brief, these authors consider the gravitational force on a parcel of gas due to the potential of the stellar disk, and show that stripping will occur if ram pressure exceeds the `restoring pressure' $P_\text{rest} = 2\pi G\, \Sigma_* \Sigma_\text{gas}$, where $\Sigma_*$ and $\Sigma_\text{gas}$ are, respectively, the stellar and gas surface densities, and $G$ is Newton's constant. To obtain the surface densities $\Sigma_*$ and $\Sigma_\text{gas}$, we divide each galaxy into a series of concentric annuli in the plane perpendicular to its velocity relative to the ICM; this does not, in general, coincide with the disk plane of the galaxy. We choose a radial bin width of 3 kpc, which is several times larger than the gravitational softening length of our simulations ($\epsilon \leq 1\, h^{-1}$ kpc). Adopting both smaller and larger widths (1 and 5 kpc, respectively) led to very similar results, as did applying a square tessellation instead. 

To calculate the ram pressure, we use the same 3000 gas particles per galaxy observation we had identified above to determine the velocity $\textbf{v}_\text{ICM}$ relative to the ICM, and also compute their mass-weighted average density $\rho_\text{ICM} = \sum_i{\rho_{\text{SPH}, i}\, m_i}/ \sum_i{m_i}$ (where $\rho_{\text{SPH},\, i}$ is the SPH density and $m_i$ the mass of the $i^{\text{th}}$ particle). The ram pressure $P_\text{ram}$ exerted on the galaxy is then simply $\rho_\text{ICM} |\textbf{v}_\text{ICM}|^2$, and the ISM in each annulus is declared `stripped' if $P_\text{ram} \geq P_\text{rest}$. In this way, we calculate the total amount of ISM expected to be (directly) stripped from each galaxy during quenching, $M_\text{direct}$, to obtain the fraction of ISM loss this can account for, $M_\text{direct}/M_\text{loss}$ (in analogy to Figs.~\ref{fig:quench.coldfate} and \ref{fig:quench.variation}).

The result is presented in Fig.~\ref{fig:quench.pram} (shaded bands, identical in the top and bottom panels); as above, we have stacked galaxies in bins of similar $z = 0$ stellar and host mass, and the band width indicates the statistical $1\sigma$ uncertainty. For ease of comparison, the \emph{total} fraction of stripped ISM is shown by dashed lines in the bottom panel (identical to the solid lines in the left panel of Fig.~\ref{fig:quench.variation}). Direct stripping clearly falls short of this, although it is far from negligible altogether --- in lower mass cluster galaxies ($\mstar \leq 10^{10}\, \msun$), for example, it accounts for $\sim 30$ per cent of ISM loss. Overall, the fraction of directly stripped ISM is approximately half the total stripped, indicating that `direct' and `fountain' stripping play an equal role in our simulations. 

\subsubsection{Stripping without feedback}
Of course, this conclusion relies on the accuracy of the \citet{Gunn_Gott_1972} ram pressure model in predicting the extent of direct stripping. As an independent test, we can exploit the fact that, as discussed in Section \ref{sec:gimic}, feedback is implemented in our simulations by giving a small number of gas particles near a newly formed star particle a velocity kick. The simulations keep track of these kicks, so that we can identify those particles which have been (directly) affected by feedback in any given snapshot interval, and likewise those which have not. We can therefore calculate the fraction of ISM loss accounted for by gas particles leaving the galaxy without having been kicked, to estimate the extent of direct stripping. 

\begin{figure}
 \centering
   \includegraphics[width=\columnwidth]{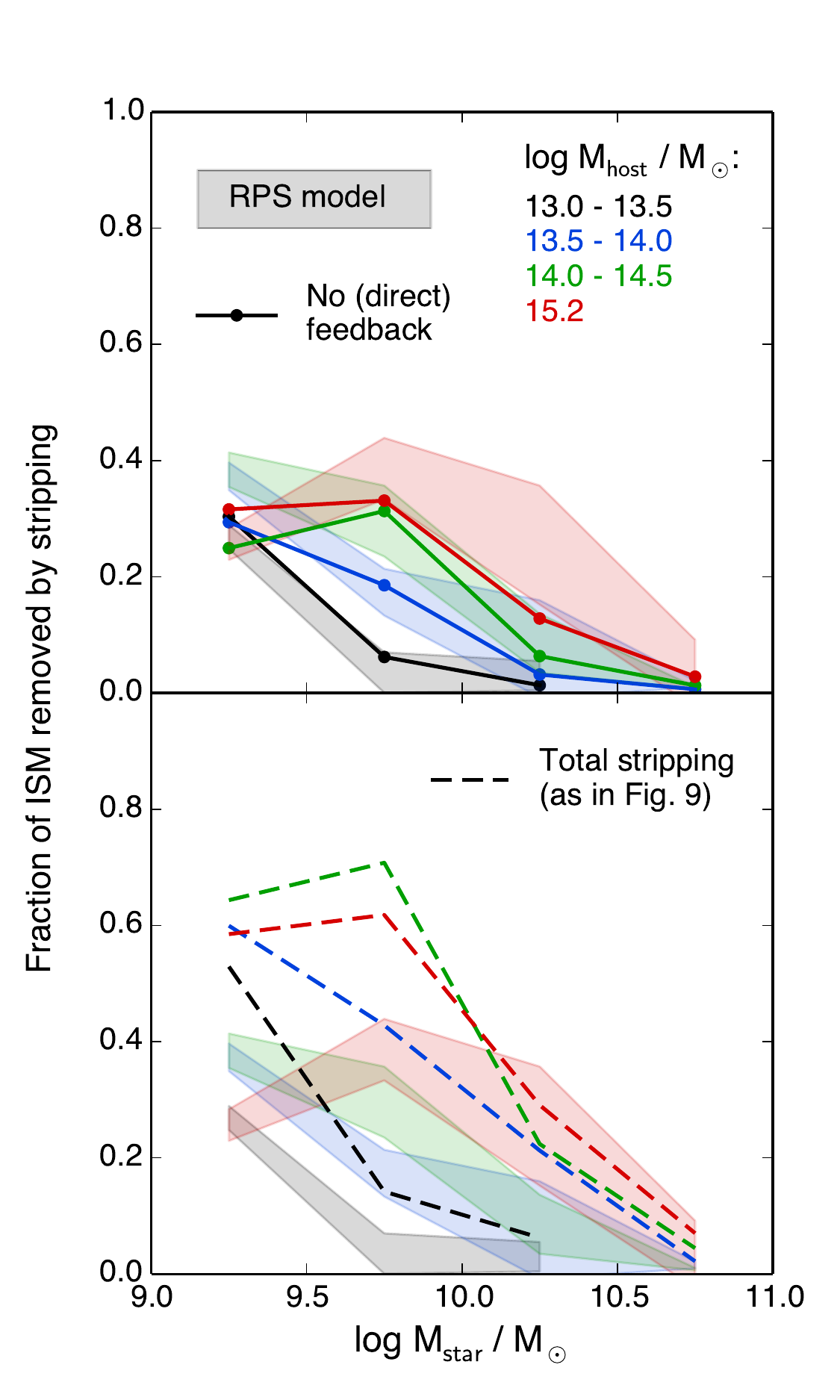}
    \caption{Comparison of the \citet{Gunn_Gott_1972} ram pressure model (shaded bands) with ISM stripping in the \gimic{} simulations. \textbf{Top:} Solid lines show the fraction of ISM that is lost from the galaxy without receiving a velocity kick from supernovae. This agrees closely with the model prediction. \textbf{Bottom}: Dashed lines give the (total) fraction of ISM that is stripped in the simulations, as shown in the left panel of Fig.~\ref{fig:quench.variation}. This is approximately twice as much as predicted by the ram pressure model, the difference being due to gas which is stripped shortly after being expelled from the star-forming disk by feedback (`stripping of fountains').}
    \label{fig:quench.pram}	
  \end{figure}

In the top panel of Fig.~\ref{fig:quench.pram}, we show this quantity with solid lines. Reassuringly, the agreement with the \citet{Gunn_Gott_1972} model (shaded bands) is remarkably good. However, the effect of feedback may well extend beyond those particles which have received direct kicks, because they can interact with other gas particles along their way and entrain these in feedback-driven winds --- note that kicked particles in \gimic{} are at no point hydrodynamically decoupled from the remaining gas \citep{DallaVecchia_Schaye_2008}. Feedback therefore affects more than just those particles directly given a velocity kick, and no part of the ISM is wholly unaffected by it (see also \citealt{Shin_Ruszkowski_2013})\footnote{In addition to the stochastic velocity kicks, the \gimic{} simulations also implement feedback through an imposed equation of state $P \propto \rho^{4/3}$ for the ISM \citep{Schaye_DallaVecchia_2008}. This prevents the formation of highly overdense regions corresponding to e.g.~Giant Molecular Clouds, and also facilitates `direct' stripping of ISM gas.}. 

Likewise, the \citet{Gunn_Gott_1972} model may overestimate the amount of gas which is stripped, because it has no possibility to account for the \emph{competing} effect of feedback and star formation: Some ISM which could be stripped by ram pressure may in fact be removed by one of these two mechanisms instead. However, the fact that there is still such a close match between the prediction of the ram pressure model and the fraction of ISM loss through not-kicked particles suggests that both effects are quite small and the importance of direct stripping in Fig.~\ref{fig:quench.pram} is not overestimated significantly. 

\subsubsection{The fate of feedback-affected particles in field and satellite galaxies}

\begin{figure}
  \centering
    \includegraphics[width=\columnwidth]{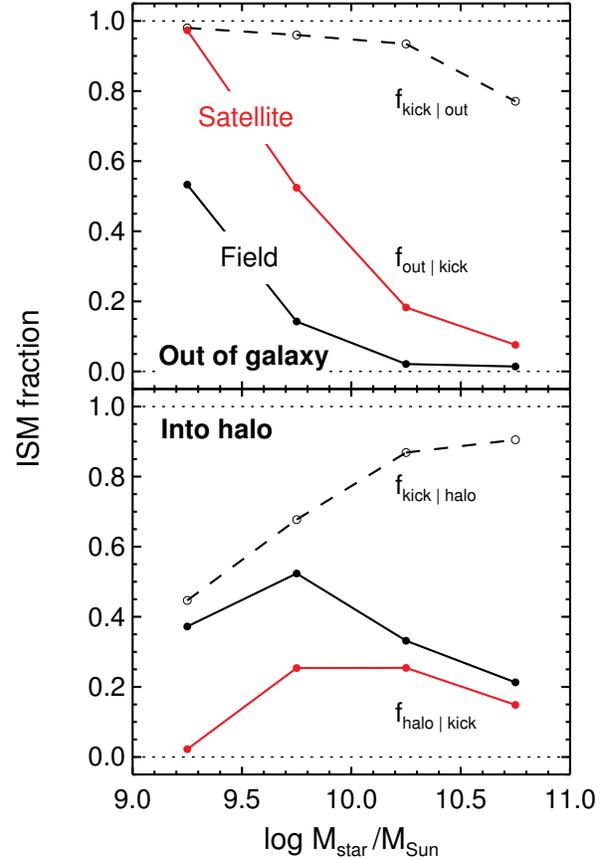}
    \caption{Effect of ISM gas being kicked by nearby supernova explosions. \textbf{Top}: Fraction that is ejected from the galaxy altogether (solid lines), in satellites during quenching (red) and equivalent galaxies in the field (black). The dashed line shows the inverse, i.e.~how much ejected gas has previously been kicked. \textbf{Bottom:} Same, but for gas recycled into the halo. Direct feedback (i.e.~particles receiving a kick) is responsible for the majority of both loss routes in the field (dashed lines). In satellites, ram pressure is then stripping part of this feedback-affected gas, leading to a higher fraction escaping the galaxy. The majority of kicked ISM particles in satellites only escape because of this `synergy' effect.}
    \label{fig:quench.fbtest}
  \end{figure}

In Fig.~\ref{fig:quench.fbtest} we confirm the action of ram pressure on ISM particles directly affected by feedback (i.e.~those given a velocity kick). Shown here is the fate of kicked particles in group and cluster galaxies during quenching (red) and the stellar-mass matched field galaxy sample (black). The top panel shows the mass fraction which is located outside the galaxy in the first snapshot after having been kicked, and the bottom panel the analogous fraction found in the halo. Even in the lowest mass galaxies ($\mstar \approx 10^9 \,\msun$), only around half the kicked particles actually escape in the field, whereas this fraction reaches almost unity in satellites. With increasing stellar mass, the `escape fraction' (top panel) decreases in both samples, but is always significantly higher in satellites than the field. This enhancement is mirrored by a a lower fraction ending up in the halo. Conversely, kicked particles make up almost the entire mass of ISM that escapes from field galaxies, and still a significant majority of gas `expelled' into the halo (dashed lines). This suggests that indirect feedback (by entrainment, rather than direct kicks) only plays a minor role, and gives further confidence that Fig.~\ref{fig:quench.pram} accurately represents the balance between direct and feedback-aided stripping in our simulations. We also note that some kicked particles remain in the ISM (or rejoin it on a timescale of $\lesssim 200$ Myr), which is why the solid lines in the top and bottom panels of Fig.~\ref{fig:quench.fbtest} do not sum to unity.

\subsubsection{Comparison to ram pressure implementation in semi-analytic models}
Our conclusion above that a substantial part of ISM stripping in the \gimic{} simulations is accomplished in synergy with feedback agrees well with results from semi-analytic galaxy formation models. \citet{Font_et_al_2008} found that the extent to which gas which has recently been `reheated' from the disk into the halo is stripped from satellites has a strong impact on the resulting passive fractions, from environmental trends being almost non-existent to much stronger than observed. These authors adopted a `stripping efficiency' factor of 0.1 to match observations (see also \citealt{Guo_et_al_2011}). Although we cannot directly identify this `reheated' component with particles receiving a velocity kick in our simulations (for instance, Fig.~\ref{fig:quench.fbtest} shows that in field galaxies typically less than half of them actually end up in the halo), the qualitative conclusion is similar: ram pressure stripping is aided by the action of supernova feedback, but not all reheated gas is removed from satellites. Note, however, that our galaxy sample was explicitly selected to have been quenched at $z = 0$, which makes it difficult to quantitatively compare our results to these prescriptions.  

An important difference to the above-mentioned models is the possibility of ISM stripping directly from the disk. As we have shown in Fig.~\ref{fig:quench.pram}, up to 40 per cent of ISM loss in the \gimic{} simulations is due to the direct effect of ram pressure stripping on ISM gas which has not been affected by feedback. This mechanism is not currently implemented in semi-analytic models, and our results suggest that it is an important --- even if not dominant --- route in which satellite galaxies lose star forming gas.

\subsubsection{Competition between feedback and stripping}
\label{sec:competition}
We have so far concentrated on the synergy between ram pressure and supernova feedback in removing ISM from group and cluster galaxies. However, both effects are also competing (together with star formation) for the removal of the same gas, which may lead to a \emph{decrease} in the effect of stripping. We can quantify the importance of this competition by examining how much of an effect stripping \emph{would} have if it was the \emph{only} mechanism affecting our satellites, and could operate past the point where the galaxies actually lose all their gas in the simulations.

\begin{figure}
  \centering
    \includegraphics[width=\columnwidth]{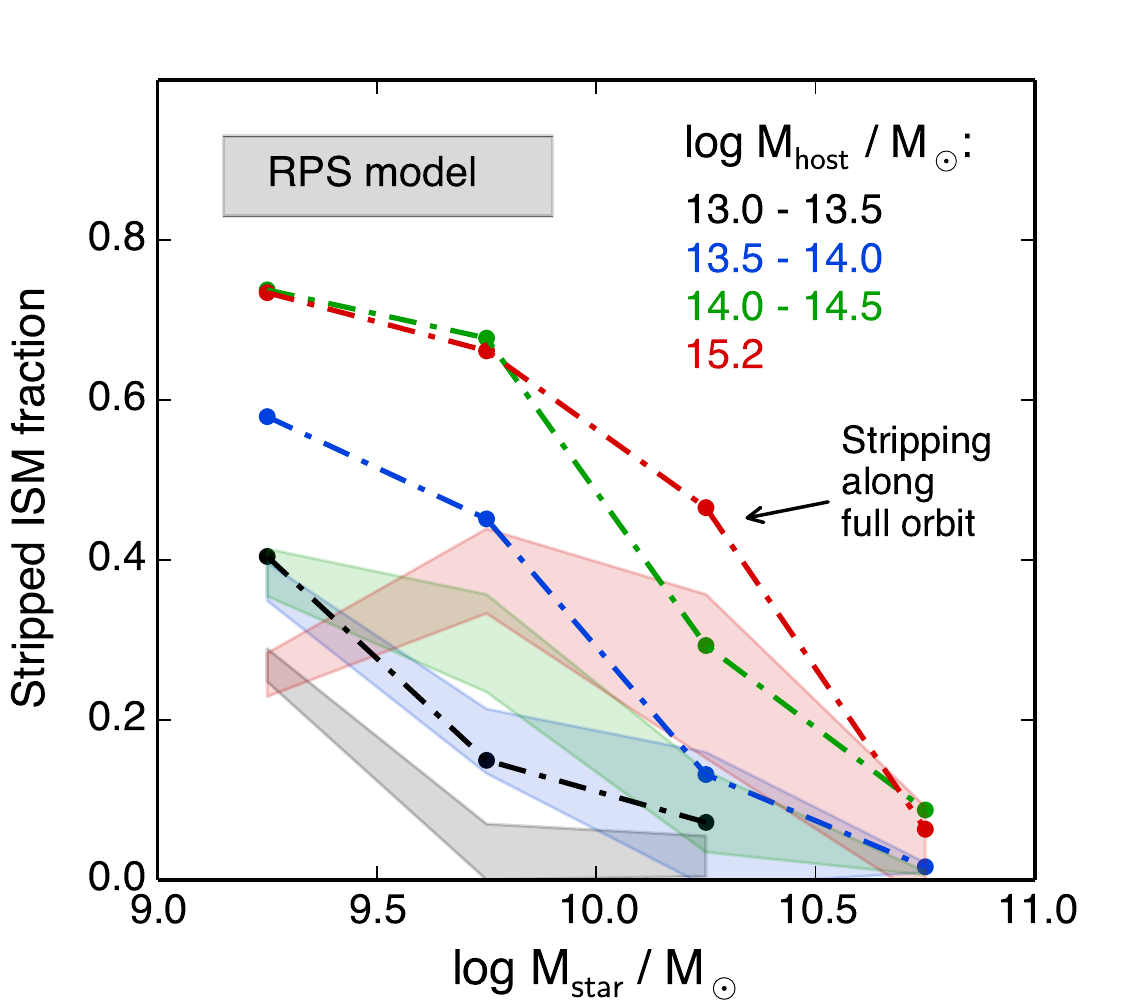}
    \caption{The \emph{hypothetical} fraction of ISM that could have been directly stripped from the disk in the absence of competing removal mechanisms (dash-dot lines), compared to how much ISM stripping is actually possible in the presence of star-formation and feedback (shaded bands). For all except the most massive galaxies in our sample, the effect of ram pressure stripping is severely reduced by this competition, which removes gas before ram pressure becomes strong enough.}    
    \label{fig:quench.eternal}
  \end{figure}

To this end, we repeat our ram pressure calculation above, but this time taking the galaxy observation from the `last normal' snapshot (with sSFR similar to the field), and calculate what fraction of this ISM could be removed by the ram pressure acting on the galaxy in each subsequent snapshot. We then find the maximum fraction of the (initial) ISM that could be stripped in this way, which is shown with the dash-dot lines in Fig.~\ref{fig:quench.eternal}. For comparison, the shaded bands show the fraction of (actual) ISM that can be stripped by ram pressure, taking into account its depletion along the galaxy's orbit (i.e.~as shown in Fig.~\ref{fig:quench.pram}).

At the high-$\mstar$ end, the difference is small: Even if there was no competition from internal processes, ram pressure would never be able to strip a substantial fraction of the ISM. The situation looks quite different for galaxies of lower stellar mass: for example, at $\mstar \approx 10^9\, \msun$ only $\sim 30$ per cent of ISM in galaxies is actually lost through stripping, but this fraction could be as high as $\sim 75$ per cent (in clusters) if ram pressure could act along the whole orbit. In fact, this is even slightly higher than the overall fraction of ISM lost through stripping, including the support of feedback (see Fig.~\ref{fig:quench.variation}). Note that this does not conflict with our earlier assertion that there is not significant competition by star formation and feedback for gas which can be stripped at a given point in time: The gas is removed (long) before ram pressure would have a chance to strip it directly. \emph{In our simulations, stripping and feedback are involved in a complex interplay, in which they enhance each other's effect, but also compete for the removal of the same gas.} 

The importance of this competition between ram pressure and feedback has also recently been highlighted in an  observational study by \citet{McGee_et_al_2014}. In close agreement to our findings, these authors concluded that the internal consumption of star-forming gas should quench satellite galaxies prior to ``orbit-based'' stripping, for example through ram pressure. However, our simulation analysis also suggests that \emph{some} gas is retained long enough for ram pressure to be effective. It is likely that galaxy-to-galaxy scatter plays an important role in explaining this discrepancy: Some galaxies may retain their ISM for longer than typical, and these may then be affected by ram pressure stripping. Similarly, the level of ram pressure experienced by different galaxies at the same group-/cluster-centric radius may vary as well, as we discuss in the following section.

\subsubsection{Summary}
To summarise our results from above: The strong enhancement of outflows seen in simulated satellite galaxies compared to the field is driven by ram pressure stripping, while contributions from tidal forces are negligible. In turn, this stripping effect is due in roughly equal parts to direct removal of gas from the disk (as considered by \citealt{Gunn_Gott_1972}) and stripping of gas almost immediately after having been expelled from the ISM into the halo. Besides this synergy effect, ram pressure also competes with feedback (and star formation) for removal of the same gas, which reduces its impact considerably, by approximately two thirds in low mass cluster galaxies.

\subsection{Reasons for importance of ram pressure}
Even after accounting for the synergy effect between ram pressure and feedback, our conclusions on the importance of ram pressure stripping appear in tension with several previous studies arguing that ram pressure stripping should only be relevant for galaxies on extreme orbits carrying them into the central regions of massive clusters (e.g.~\citealt{Gunn_Gott_1972,Abadi_et_al_1999}). We now discuss two potential reasons for the larger-than-expected influence of ram pressure: scatter in the ram pressure profile of groups and clusters, and a varying importance of stripping over cosmic time.

\subsubsection{Ram pressure scatter}
\begin{figure}
  \centering
    \includegraphics[width=\columnwidth]{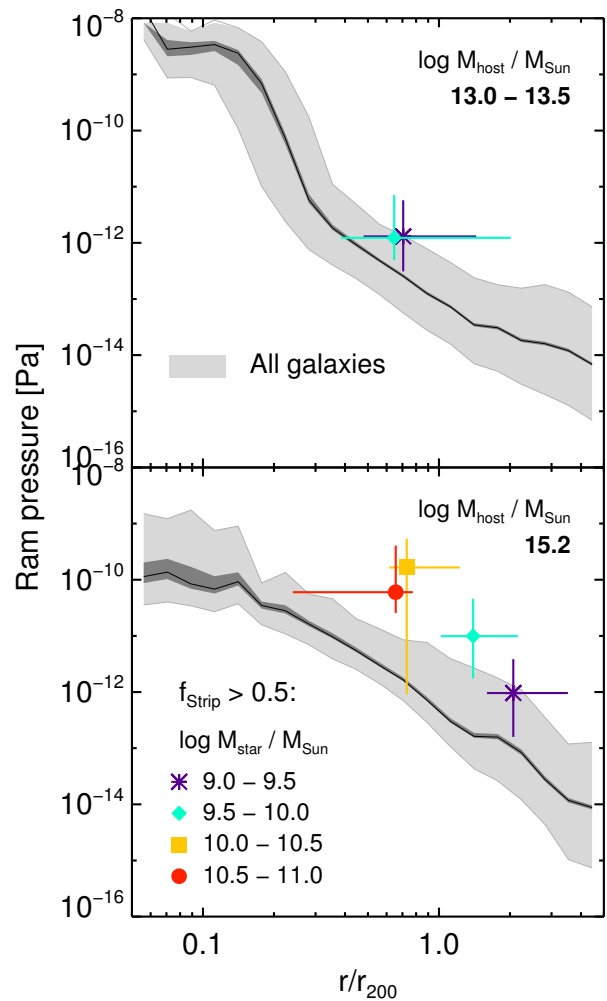}
    \caption{Ram pressure profile for all galaxies (grey) in low-mass groups (\textbf{top}) and the massive cluster (\textbf{bottom}). Overplotted in colour are the median radius and ram pressure of galaxies undergoing significant stripping ($f_\text{strip} \geq 0.5$). This occurs preferentially in regions of unusually high ram pressure, and is not restricted to the central group/cluster regions.}
    \label{fig:quench.rpprofile}
  \end{figure}

We first explore how much the ram pressure experienced by our simulated group and cluster satellites varies at a given radius. In an independent simulation, \citet{Tonnesen_Bryan_2008} had found that this variation could be as large as one order of magnitude in a massive cluster, so that individual galaxies can experience much stronger forces than what is typical at a given radius. In Fig.~\ref{fig:quench.rpprofile}, we show the ram pressure profiles of host haloes in the \gimic{} simulations. For simplicity, we here only analyse the two most extreme halo mass bins, and show a stack of our low-mass groups ($13.0 \leq \log\, M/\,\msun \leq 13.5$) in the top panel, and the massive cluster ($\log\,M/\,\msun \approx 15.2$) in the bottom. The black line represents the median ram pressure experienced by all galaxies at a given radius. The dark grey band indicates its statistical $1\sigma$ uncertainty: $\sigma = P_{50} + (P_{15.9/84.1} - P_{50}) / \sqrt{N}$, where $P_n$ is the $n^\text{th}$ percentile of the ram pressure distribution -- so that $P_{50}$ corresponds to the median -- and $N$ the number of galaxies per bin. The light grey band indicates the extent of galaxy-to-galaxy scatter, by showing the $25^\text{th}$ and $75^\text{th}$ percentiles. In addition, the figure also shows the median $r/r_{200}$ and $P_\text{ram}$ of galaxies undergoing stripping ($\gtrsim 50$ per cent of instantaneous ISM loss through ram pressure), different colours corresponding to different stellar masses. Statistical $1\sigma$ uncertainties are here indicated by the corresponding errorbars. 

In both host mass bins, there is significant scatter around the median ram pressure profile, with the 25$^\text{th}$ and 75$^\text{th}$ percentiles separated by almost two orders of magnitude. Naturally, high ram pressure surroundings favour stripping, so it is perhaps not too surprising that galaxies which are actually being stripped are found at the very top end of this distribution, with a positive bias in ram pressure of approximately one order of magnitude in the case of low-mass groups and up to two in the massive cluster, compared to what is typical at a given radius. In Appendix \ref{sec:app.rpscatter}, we show that this bias is mostly due to these galaxies being located in highly overdense regions of the ICM typical of filaments (see also \citealt{Bahe_et_al_2013}). This agrees well with \citet{Tonnesen_Bryan_2008}, who also concluded that ICM substructure could expose galaxies to strongly enhanced levels of ram pressure, and explains at least in part why our simulations predict ram pressure stripping to be much more common than what would be expected from the `typical' levels (e.g.~\citealt{Gunn_Gott_1972,Abadi_et_al_1999}). 

There are two further features of Fig.~\ref{fig:quench.rpprofile} that are worth pointing out explicitly. Firstly, stripping is, in our simulations, evidently not restricted to the central regions of either groups or clusters, and instead mostly occurs in the outskirts, at $r \gtrsim 0.5\, r_{200}$. Galaxies with $\mstar < 10^{10}\, \msun$ near the massive cluster are even stripped significantly outside $r_{200}$. This means that it is unlikely that overcooling of the central group and cluster regions, and the associated unphysically high ICM densities near the centre, have a significant impact on our results --- in the case of the low-mass groups, for example, there is a noticeable upturn in ram pressure only at $r \lesssim 0.2\, r_{200}$. Secondly, there is no pronounced discontinuity in the ram pressure profiles around $r_{200}$, so that it is not too surprising to find galaxies being stripped outside this radius as well. As we show in Appendix \ref{sec:app.rpscatter}, there are actually pronounced changes in the profiles of both ICM density and velocity around this radius, but in opposite directions so that they largely cancel out in the resulting ram pressure profile.

\subsubsection{Quenching epoch}
\begin{figure}
  \centering
    \includegraphics[width=\columnwidth]{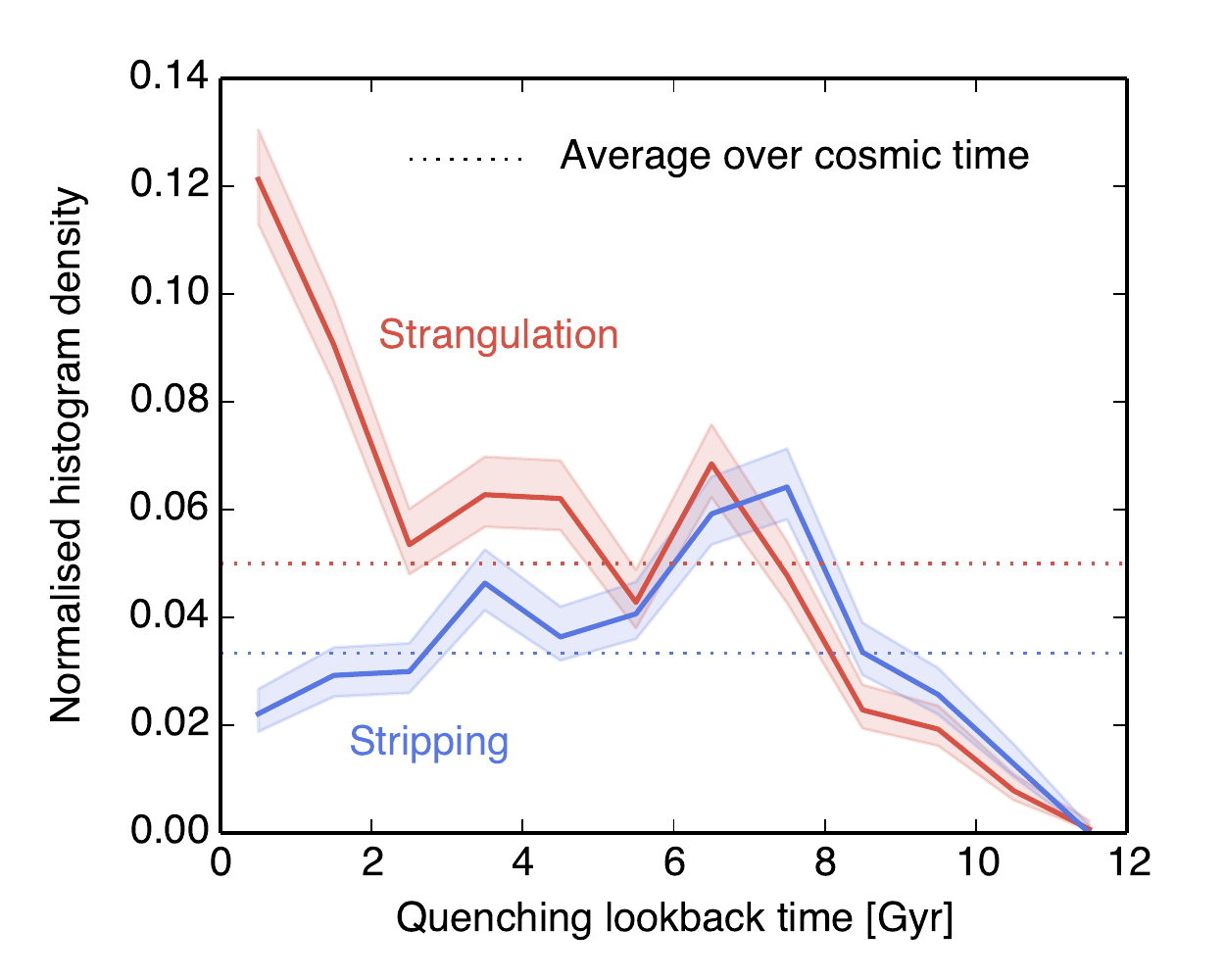}
    \caption{Redshift evolution of stripping and strangulation for $z = 0$ quenched galaxies. Shown is the distribution of lookback times to the point of quenching, for `stripped' and `strangled' galaxies (blue and red, respectively). Stripped (strangled) have lost more (less) than 50 per cent of their ISM due to ram pressure stripping, direct or from fountains. Both distributions are normalised to the total number of quenched galaxies. The ratio between galaxies quenched through stripping and strangulation at late times ($\sim$ 1:6) is much lower than integrated over cosmic time ($\sim$ 1:1.5).}
    \label{fig:quench.acctime}
  \end{figure}

Another key characteristic that one may expect to be related to the mechanism of ISM removal is the \emph{epoch} when the galaxy was quenched, whose role we explore in Fig.~\ref{fig:quench.acctime}. We show here the normalised distribution of lookback times to the last snapshot in which each galaxy was star-forming (sSFR $\geq 10^{-11}$ yr$^{-1}$), separated into galaxies whose ISM loss is dominated by stripping (`stripped', blue) or internal processes (`strangled', red). If all galaxies had been quenched uniformly throughout cosmic history, both would be at the level of the horizontal dotted lines, but in reality, this is clearly not the case. In particular, the distributions of strangled and stripped galaxies differ significantly: strangulation has become progressively more common towards later times (lower lookback time), whereas stripping had a peak around eight Gyr ago ($z \approx 1$) and has since declined in importance. By construction, both samples only include galaxies which still exist (i.e.~are not completely disrupted) at $z = 0$, and it is well possible --- and indeed expected --- that a number of galaxies which were quenched at higher redshift have since been disrupted. This may explain the increased importance of strangulation at low redshift, but not the decline in stripping activity: \emph{the \gimic{} simulations predict that stripping was more common for galaxies accreted at higher redshift.} 

We speculate that this effect may be due to the higher sSFR in the past (and associated stronger fountains), or an increased density of the ICM at higher redshift. In a future paper, we will investigate the physical origin of this trend in detail. For simplicity, we have here combined all quenched satellite galaxies, but there is no significant difference when analysing only a subset of our sample with a narrower range of $\mstar$ or $M_\text{host}$. 

One important consequence is that observations of galaxies being quenched at the present epoch will significantly underestimate the importance of stripping in building up today's population of passive group and cluster galaxies. The ratio of stripping to strangulation at $z \approx 0$ is $\sim$ 1:6, which is around four times lower than its average over cosmic history ($\sim$ 1:1.5). We speculate that this may be another key reason why ram pressure has so far been mostly discarded as a viable quenching mechanism, in addition to the other effects we have discussed above (synergy between stripping and feedback, and locally enhanced ram pressure due to filaments).


\section{The reason for the lack of ISM replenishment}
We have so far concentrated on understanding why the ISM which is already present in group and cluster galaxies is removed. However, to quench star formation it is also necessary to prevent further ISM replenishment, the cause of which we analyse in this section. 

\subsection{Depletion of the gaseous halo prior to quenching} 

\begin{figure}
  \centering
    \includegraphics[width=\columnwidth]{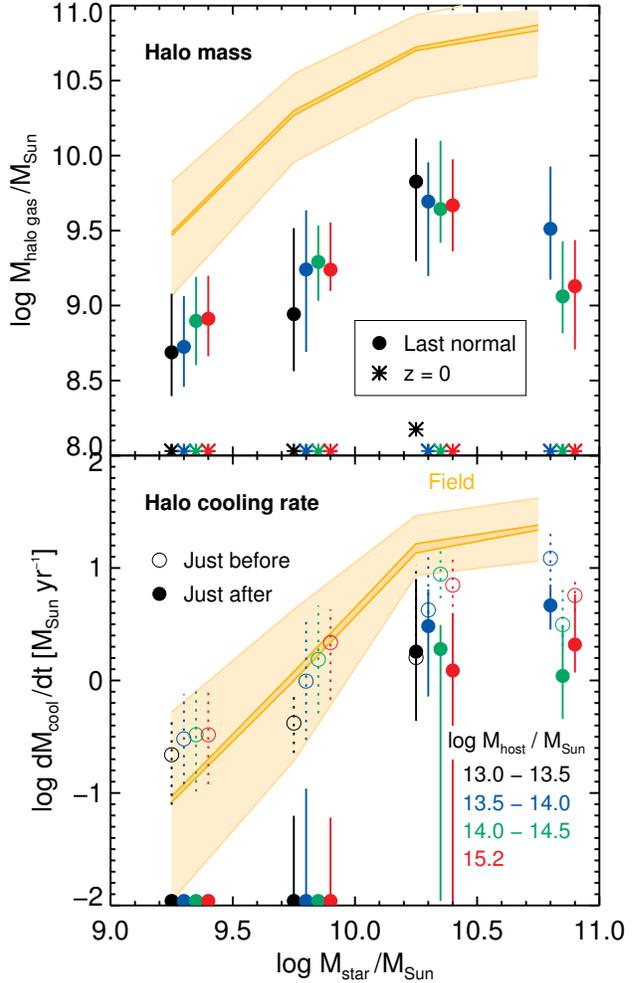}
    \caption{\textbf{Top:} Amount of halo gas in galaxies at their last `normal' snapshot (when their sSFR was last consistent with the field; filled circles) and at $z = 0$ (asterisks). \textbf{Bottom:} The rate at which the ISM is replenished from this gaseous halo, immediately before (open) and after (filled) this point. Even before star formation is affected, galaxies are already depleted in halo gas, which explains the significant lack of ISM replenishment evident from the bottom panel. Note that values below the plot range are shown at the bottom end of each panel for clarity.}
    \label{fig:quench.lastnormal}
  \end{figure}

 In Fig.~\ref{fig:quench.lastnormal}, we investigate to what extent ISM replenishment from the galaxy halo is suppressed in satellites. The top panel shows the mass of `halo' gas (i.e.~that which is not part of the ISM; note that this includes both hot gas and some cold gas which is not dense enough to be forming stars) in the last normal snapshot, when the sSFR was last consistent with the field. Filled circles represent the median mass in bins of equal stellar and host mass, with 25$^\text{th}$ and 75$^\text{th}$ percentiles indicated by vertical lines. For clarity, symbols representing different host masses have been offset slightly from each other. The dark yellow band indicates the median halo gas mass in the field (its width representing statistical uncertainty), and the light yellow band the $1\sigma$ scatter.

It is immediately clear that, even though star formation is by definition still unaffected by the environment at this point, the gaseous halo has already been depleted significantly. The depletion is strongest for the most massive galaxies (almost two orders of magnitude difference in halo gas mass), but still almost one order of magnitude for galaxies with $\mstar \approx 10^9\, \msun$. There is no significant systematic influence of host mass: at the onset of star formation quenching, galaxies in a massive cluster have their halo depleted similarly as those in a poor group. We emphasise however that this sample only includes galaxies quenched at $z = 0$, which make up a much larger fraction of the overall population in more massive hosts (see Fig.~\ref{fig:quench.ssfrspec}). The asterisk symbols in the top panel of Fig.~\ref{fig:quench.lastnormal} confirm that almost all galaxies in our sample have completely lost their gaseous haloes by $z = 0$.

In the bottom panel, we verify that this halo gas depletion has a dramatic effect on the replenishment of the ISM. In a similar way to the top, we show here the rate at which gas cools and condenses from the halo into the ISM, $\dot{M}_\text{cool}$. Because computation of this rate involves comparing two snapshots, we show both the rate obtained over the snapshot interval immediately before (empty circles) and after (filled) the last normally star forming snapshot. Replenishment is severely reduced --- in many cases to zero --- in the latter case, which is a direct consequence of the reduction in halo gas mass. Interestingly, this drop in replenishment happens very quickly for galaxies with $\mstar \lesssim 10^{10}\, \msun$: over the preceding snapshot interval, it was still consistent with the field.

We conclude that ISM replenishment is shut off (or at least severely reduced, in the case of galaxies with $\mstar \gtrsim 10^{10}\, \msun$) before quenching begins. The reason why star formation is then not beginning to be quenched immediately is likely that it is dominated by the densest ISM regions, which are themselves replenished from the less dense parts of the ISM, so that these provide a `buffer', albeit only for a limited time. Note that, in principle, it would also be possible for the ISM to be directly replenished from outside the galaxy (for example, through mergers), however we have checked for this and found no significant contribution from this channel (see also \citealt{Feldmann_et_al_2011} and \citealt{Behroozi_et_al_2014}).

\subsection{Halo depletion mechanisms}

As a final piece of our investigation, we can now ask how the gaseous halo has been removed. By definition, the halo gas is not dense enough to directly form stars, so that it can only be lost through outflows or cooling. We separate these in analogy to our analysis of the ISM loss in Section \ref{sec:ismanalysis}. Recall that this requires determining for each galaxy a starting point, from which on we sum up the mass of gas lost in the different ways. Above, we had chosen the time when the galaxy's sSFR started deviating from the field, but this is not appropriate here: As shown in Fig.~\ref{fig:quench.lastnormal}, the halo is already strongly depleted in this `last sSFR-normal' snapshot. We therefore define an analogous starting point for our present purpose as the time when the \emph{mass of the gaseous halo} was last consistent with the field, i.e. $$ M_{\text{gas halo}}^{\text{sat}} \geq M_{\text{gas halo}}^{\text{field}} - \sigma (M_{\text{gas halo}}^{\text{field}})$$ where the last quantity is the scatter in gas halo masses within the comparison field galaxy sample (see Section \ref{sec:onsetdetermination}). 

\subsubsection{Relative importance of stripping, feedback-driven outflows, and cooling}

\begin{figure*}
  \centering
    \includegraphics[width=2.1\columnwidth]{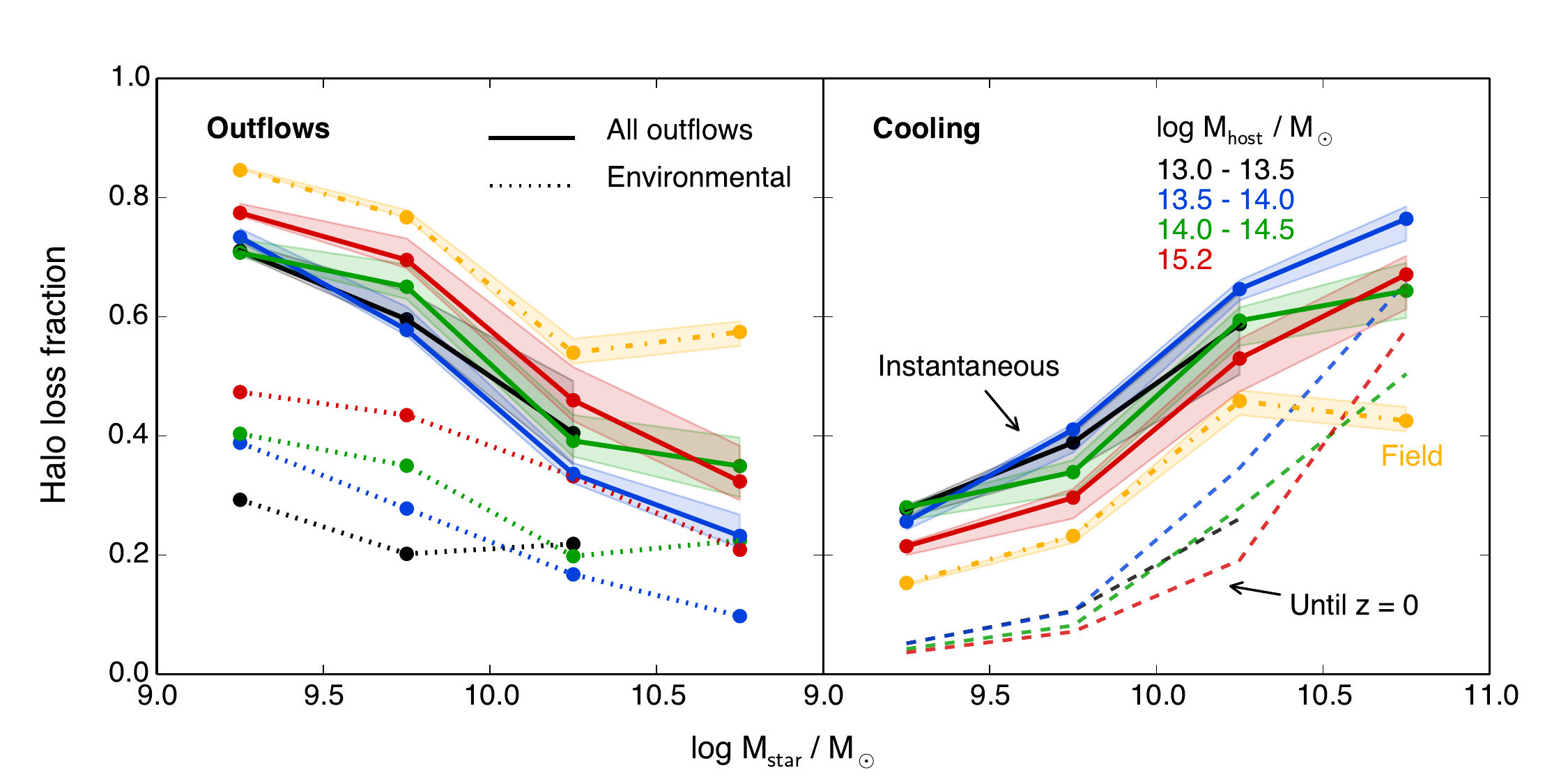}
    \caption{The mechanisms responsible for removing the gaseous halo in group and cluster galaxies. The \textbf{left} panel shows the fraction of loss attributed to outflows of gas, the \textbf{right} one due to cooling (i.e. replenishment of the ISM). The dotted line in the left panel gives the fractional loss due to environment alone (see text), which is considerably lower than the total outflow loss. In contrast to the loss of ISM, the host mass plays only a minor role here, but there is a significant trend with stellar mass: outflows dominate at low stellar masses, cooling at the high-mass end. Note that, although the \emph{fractional} loss through cooling and outflows is similar to the field (yellow), the absolute loss is much lower (see Fig.~\ref{fig:quench.lastnormal}).}
    \label{fig:quench.halofate}
 \end{figure*}

Fig.~\ref{fig:quench.halofate} shows the fraction of halo gas lost through outflows in the left panel, and on the right that lost through cooling. To aid interpretation of the outflow part, we also show the satellite-specific part of these outflows with dotted lines in the left panel, i.e.~the excess compared to the field calculated as described in Section \ref{sec:envexc}. 

As with the ISM, there is clearly no one process that dominates the loss of halo gas across all bins of stellar and host mass. One pronounced difference, however, is the apparently small role of host mass on the fraction of outflows: there are hardly any cases where the result differs by more than $\sim 10$ per cent between the two most extreme cases, the massive cluster and low-mass groups (red and black, respectively). Perhaps even more surprising is that the satellite-specific outflow part (i.e.~what is caused by stripping) is by no means dominating the removal of the halo: At the low-$\mstar$ end, it accounts for only $\sim 50$ per cent of its loss, dropping to $\sim 20$ per cent at $\mstar \approx 10^{11}\, \msun$ even in the massive cluster.

A comparatively strong loss channel, on the other hand, is that of cooling, i.e. halo gas becoming part of the ISM and potentially turning into stars later. For galaxies with $\mstar \gtrsim 10^{10}\, \msun$ this accounts for the majority of halo loss --- although the lack of AGN feedback in our simulations may mean that this increase is an overestimate --- and is still quite significant at the low mass end ($\sim 20$ per cent). Although the \emph{direct} effect of SN feedback is mostly limited to the ISM, gas in the halo may nevertheless be entrained in these `primary' winds (see also Fig.~\ref{fig:quench.fbtest}). The fraction of outflows not accounted for by stripping is therefore most likely due to this indirect feedback effect\footnote{Another possibility is that halo gas cools to the ISM and is almost immediately kicked by a nearby star particle, but this does not occur commonly enough in our simulations.}.

A comparison to the equivalent loss fractions in field galaxies (yellow) reveals that --- perhaps surprisingly --- the \emph{fraction} of halo gas which cools into the ISM is remarkably unaffected by the environment, and is in fact slightly enhanced in groups and poor clusters. One way to interpret this is that gas which is stripped --- and this fraction does increase with increasing host mass --- is predominantly that in the outer parts of the halo where the cooling time is longest, so that it would not have been (directly) relevant to the replenishment of the ISM anyway. The gas which is actually cooling, and therefore of relevance to the future ability of the galaxy to form stars, on the other hand, is largely left in peace by stripping, which would explain why cooling is --- in relative terms --- slightly more important than in the field. Furthermore, the fact that there is a (slight) trend with host mass in the cooling fraction which is \emph{opposite} to the offset between the field and satellites in general, would imply that stripping has a slightly more direct impact on the `cooling' part of the halo in a massive cluster than in groups, for example because its higher levels of ram pressure are able to remove halo gas down to smaller galacto-centric radii.

\subsubsection{Halo (non-)replenishment from cosmological accretion}
\begin{figure}
\centering
    \includegraphics[width=\columnwidth]{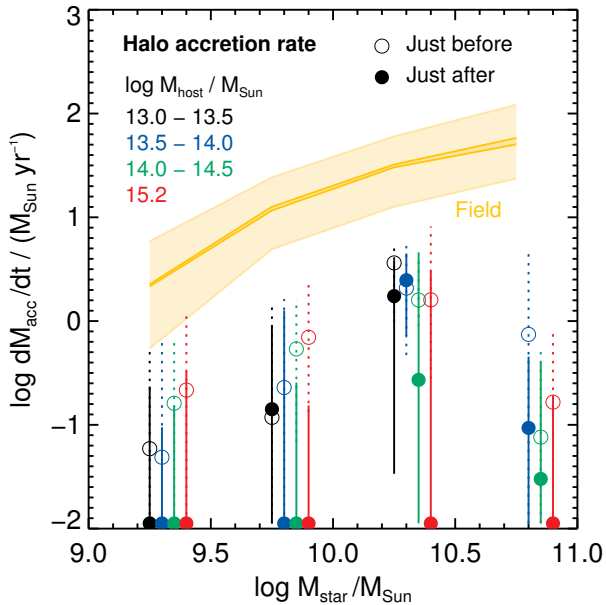}
    \caption{Rate of cosmological accretion onto the gaseous halo in field galaxies (yellow) and satellites in groups and clusters, at the onset of star formation quenching (coloured circles; open ones show the rate immediately before this point, filled circles immediately after). The accretion rate is substantially suppressed in the group/cluster sample, and drops to zero after the last normal snapshot for the majority of galaxies (filled circles). This lack of cosmological accretion is a key driver of the depletion of the gaseous halo.}
    \label{fig:quench.haloaccretion}
  \end{figure}

The trends in Fig.~\ref{fig:quench.halofate} therefore imply that the halo itself is largely depleted by a `strangulation' effect: the majority of its gas is used up internally through cooling and entrainment in winds, with the key difference to field galaxies being a much reduced rate of cosmological gas accretion. We confirm this last point explicitly in Fig.~\ref{fig:quench.haloaccretion}, where we compare the rate of cosmological gas accretion onto the halo in the field (yellow) and satellite galaxies at the last (sSFR-)normal snapshot, in analogy to Fig.~\ref{fig:quench.lastnormal}. There is a clear discrepancy of more than an order of magnitude, and immediately after this point, cosmological accretion drops to zero in the majority of galaxies. The importance of this reduction in cosmological accretion has also previously been stressed by \citet{Feldmann_et_al_2011} in the case of galaxy groups, a conclusion that we both confirm and extend to all halo masses above $10^{13} \msun$.

One caveat with the above analysis is that it is based on the \emph{instantaneous} loss fractions. Particles which have cooled into the ISM can be lost from the galaxy through outflows later on, so that some of the initially `cooled' gas can later be stripped. To see the influence of this `delayed halo stripping', we have repeated our analysis, this time looking at the fate of the particles at $z = 0$ (instead of the first snapshot when they were no longer in the halo, as above). This is shown with the dashed lines in the right panel of Fig.~\ref{fig:quench.halofate}. At the high-$\mstar$ end, the difference is very small, but at lower stellar masses, the latter analysis gives a noticeably lower cooled fraction, which drops to only $\sim 5$ per cent at $\mstar \approx 10^9 \msun$. This is easily understood in terms of the increased importance of outflows in carrying away ISM from lower mass galaxies, as evident from Fig.~\ref{fig:quench.coldfate}. 

The simulations therefore predict a rather complex picture for the removal of the gaseous halo: One part is removed directly by stripping and entrainment in galactic winds, while a significant fraction initially cools to the ISM, some of which is \emph{later} removed from there (for example, when the galaxy has moved closer to the group/cluster centre and is subject to stronger levels of ram pressure).  We finally note that there is no contradiction between the \emph{fraction} of the halo removed through cooling not being lower than in the field and the drastically reduced actual halo cooling rate as shown in Fig.~\ref{fig:quench.lastnormal}: it simply means that the overall rate of halo loss is (much) lower than in the field.

\subsubsection{Comparison to idealised simulations}
One direct consequence of this mix of halo depletion mechanisms is that the result of \citet{McCarthy_et_al_2008b} --- who showed that ram pressure \emph{alone} leads to a rather gradual stripping of a galaxy's (hot) halo with some fraction ($\sim 30$ per cent) able to remain intact even many Gyr after accretion --- is only a relatively weak lower limit on the actual depletion of halo gas in satellite galaxies. On the other hand, however, the influence of halo stripping on the galaxy's star formation rate may be much less severe than on the mass of the halo itself. This could have potentially important implications for semi-analytic models of galaxy evolution in a group or cluster, which have traditionally had difficulties in reproducing the observed passive fraction of satellite galaxies (see e.g.~\citealt{Weinmann_et_al_2006, Font_et_al_2008, Guo_et_al_2011, Henriques_et_al_2013}).

\section{Discussion and Summary}
\label{sec:discussion}

We have studied star formation quenching in a large sample of group and cluster galaxies drawn from the \gimic{} suite of cosmological hydrodynamical simulations. The aim of this work has been to identify the main physical mechanism(s) responsible for shutting down star formation in dense environments, as well as to understand the reason for their importance. Our main conclusions may be summarised as follows:

\begin{enumerate}

\item The \gimic{} simulations reproduce the observational result of a bimodal sSFR distribution, with an active (star-forming) peak whose position does not vary significantly with environment. This is explained by rapid quenching (in particular for low-mass galaxies, $\mstar \lesssim 10^{10}\, \msun$), and more recent accretion of galaxies still forming stars at $z = 0$ compared to those which are passive, by typically $\sim 2$ Gyr.

\item There is a large degree of scatter in when galaxies become passive, ranging from $\sim 2$ Gyr prior to accretion (first crossing of $r_{200,\,c}$) to $> 4$ Gyr after. Quenching generally occurs earlier for less massive galaxies, and those orbiting massive hosts.

\item Outflows are a significant contributor to star formation quenching in the \gimic{} simulations, accounting for up to $80$ per cent of ISM removal in cluster galaxies with $\mstar \lesssim 10^{10}\, \msun$; their exact role depends on both galaxy and host halo mass. Direct star formation only removes $\sim 50$ per cent of the ISM even in the most massive galaxies we have studied, and many of these would already have become passive of their own accord in the real Universe. 

\item Outflows in our simulations are driven by both internal processes (feedback) and external stripping (ram pressure only, we find no evidence for a substantial contribution from tidal forces). Approximately half the stripped ISM has also been affected by feedback in a `stripping of galactic fountains' scenario, while the other half is stripped directly from the star-forming disk.

\item Stripping is more relevant just before star formation stops than integrated over the whole period of quenching. This suggests an `interrupted strangulation' scenario, where gas loss is initially dominated by internal consumption mechanisms (strangulation), but before the ISM is completely depleted in this way, stripping becomes strong enough to complete the process.

\item Galaxies at the same radial distance from the group/cluster centre can experience a wide range of ram pressure levels (varying by an order of magnitude or more), and those that are actually stripped are preferentially found at the high end of this distribution. Stripping is not limited to the central group and cluster regions, but mostly occurs outside $\sim 0.5\, r_{200}$. 

\item For quenched satellites surviving until $z = 0$, ram pressure stripping was more effective at $z \approx 1$ than the present epoch by a factor of $\sim 3$. As a consequence, the ratio of galaxies quenched due to strangulation and due to stripping is $\sim 4$ times higher at $z \approx 0$ than averaged over cosmic history.

\item We confirm that the effect of ram pressure is well described by the model of \citet{Gunn_Gott_1972}, when only accounting for the `pure' stripping effect without influence of feedback. Application of this model to the full galaxy orbits has shown that at the low-mass end ($\mstar \approx 10^9\, \msun$) stripping \emph{could} be up to three times as effective (removing up to 75 per cent of the ISM) if it were not for the competing effects of star formation and feedback.

\item ISM replenishment is suppressed because the gaseous halo surrounding the star forming disc is removed before star formation is quenched. This removal is not exclusively a consequence of ram pressure (only to $\lesssim 50$ per cent even in the massive cluster), with internally driven process (cooling and indirect feedback) being of relevance as well. The shutdown of cosmological accretion is therefore another physical mechanism co-responsible for star formation quenching in group and cluster galaxies.
	
\end{enumerate}

The median evolution of our simulated $z=0$ quenched galaxies as they are accreted onto a group or cluster halo is summarised in Table \ref{tab:evolution}.  At all ranges of stellar mass and host mass that we consider, cosmological gas accretion from the field ceases well before the galaxy crosses $r_{200}$ for the first time. As a result, the gaseous halo surrounding the galaxy is depleted at accretion to only $\lesssim 20$ per cent of the halo mass in comparable field galaxies. The star formation rate typically drops below the $1\sigma$ range of field galaxies from $\sim$ 1 Gyr before accretion (in the case of the lowest mass galaxies falling into the most massive cluster) to  $\sim$ 1.5 Gyr after for the opposite extreme (Milky-Way analogues in groups). As a result, the sSFR is not reduced as severely at the point of accretion as the mass of the gaseous halo, and in fact slightly exceeds the field value in the case of galaxy groups. Star formation in lower mass galaxies with $\log_{10}\, (\mstar/\msun) = [9.0 - 10.0]$ is then typically quenched within 500 Myr of accretion, with $\geq 75$ per cent of star forming gas lost through stripping. In contrast, more massive galaxies are quenched only after up to 5 Gyr, with stripping accounting for $\lesssim 20$ per cent of their ISM loss: for these, internal gas consumption combined with the reduced cosmological replenishment is the dominant cause of star formation quenching.

\begin{table*}

\caption{Summary of the evolution of low- and high-mass galaxies falling into group and cluster haloes (for simplicity, we have combined all `galaxy group' scale haloes into a single bin spanning a full decade in halo mass). Given are the median times relative to accretion (first crossing of $r_{200}$) of three key events: cosmological gas accretion drops below the $1\sigma$ range of comparable field galaxies, the analogous point for sSFR, and sSFR drops below $10^{-11}$ yr$^{-1}$ so the galaxy is `quenched'. Also shown are the remaining halo mass and sSFR at accretion relative to stellar mass matched field galaxies as well as the fraction of star-forming gas lost to stripping.}

\begin{tabular}{llcccccccc} \hline \hline
\textrm{Stellar mass} & log$_{10}\, (\mstar / \msun)$  &\hspace{0.5cm} & & 9.0 - 10.0 &  &  \hspace{0.5cm} & & 10.0 - 11.0 &   \\
\textrm{Host mass} & log$_{10}\, (M_\text{host} / \msun)$ & & 13.0 - 14.0 & 14.0 - 14.5 & 15.2 & & 13.0 - 14.0 & 14.0 - 14.5 & 15.2 \\ \hline 

Gas accretion below field & [Gyr] & &-1.35 & -1.28 & -2.08 & & -0.62 & -0.93 & -0.76 \\
Halo mass at accretion & [relative to field] && 14\% & 7\% & 0\% & & 22\% & 18\% & 10\% \\
sSFR at accretion & [relative to field] && 113\% & 54\% & 0\%& & 106\% & 90\% & 94\% \\
sSFR below field & [Gyr] && 0.08 & -0.22 & -1.00 && 1.56 & 0.56 & 0.34 \\
sSFR $< 10^{-11}$ yr$^{-1}$ & [Gyr] && 0.49 & 0.17 & -0.66 && 5.00 & 4.47 & 1.80 \\
Stripped fraction of SF gas & && 75\% & 88\% & 90\% && 13\% & 12\% & 20\% \\ \hline

\end{tabular}
\label{tab:evolution}
\end{table*}

These results present important new insight into the evolution of group and cluster galaxies. In particular, the prediction that ram pressure stripping (in conjunction with feedback) should be important in hosts of all masses, out to the virial radius and beyond, means that it is not a rare process operating only in extreme circumstances, and could e.g.~be well compatible with a small intrinsic scatter in the red galaxy fraction as determined by \citet{Balogh_McGee_2010}. It also agrees with the results of \citet{Fabello_et_al_2012}, who found that strangulation alone was insufficient to explain the H\textsc{i} properties of galaxies in hosts with masses as low as $10^{13}\, \msun$.

In principle, the predictions presented here can be tested by $\textsc{Hi}$ observations of galaxies in groups and clusters, which would confirm (or rule out) the wide-spread presence of tails from ram pressure stripped gas behind infalling galaxies. However, our analysis also suggests that the majority of present-day passive galaxies were quenched as a result of stripping which occurred several Gyr ago. Such high redshifts are beyond the range of current radio telescopes, but the upcoming \emph{Square Kilometre Array} (SKA) promises to enable such a direct observational test. 

In addition, observations have already revealed that the environmental influence on group and cluster galaxies depends strongly on their surface density, as well as stellar mass \citep{Zhang_et_al_2013}. Unfortunately, the resolution of our simulations is not high enough to reliably investigate this influence --- the gravitational softening length of $\epsilon =$ 1 kpc at low z is an appreciable fraction of the typical stellar half-mass radius --- but we intend to study this relation with improved simulations in the future.

\subsection{Comparison to Wetzel et al. (2013)}
\begin{figure}
  \centering
    \includegraphics[width=\columnwidth]{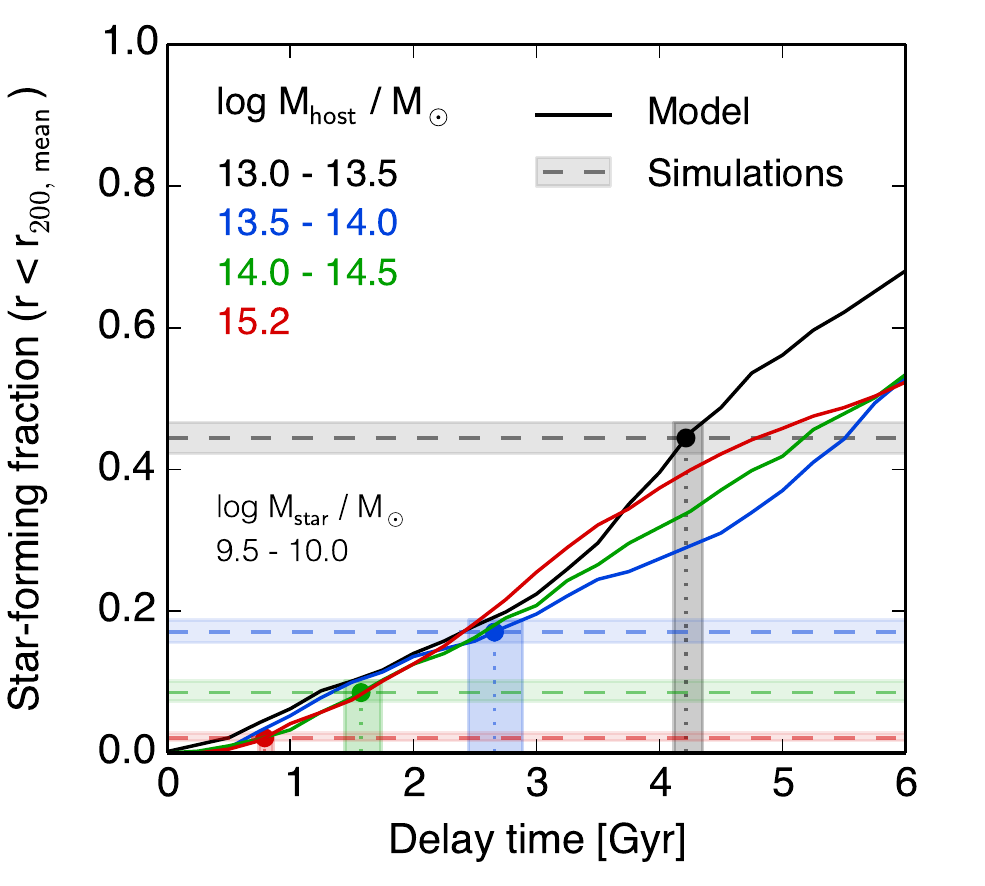}
       \caption{Fraction of star-forming $z \leq 0.2$ galaxies at $r < r_{200,\, m}$ obtained from a simple evolutionary model which assumes an initial delay time (x-axis), after which star-formation is quenched rapidly (solid lines; different colours represent different host masses as detailed in the top-left corner). Comparing these to the actual star-forming fraction of our simulations (dashed lines), one obtains the best-fit delay time (filled circles and vertical dotted lines). This is short for the massive cluster (red), but for all other host masses a relatively long delay time of $\sim 2 - 4$ Gyr is inferred.}
    \label{fig:quench.delaymodel}
  \end{figure}

Using simple merger tree based accretion models, \citet{Wetzel_et_al_2013} infer a long delay of $t_\text{delay} \approx 3$ Gyr between accretion and the onset of quenching, in order to reproduce the sSFR distribution of SDSS satellites. By contrast, we have found a wide range of timescales (from $\sim 2$ Gyr prior to accretion up to $> 4$ Gyr after). In order to compare our results in a fair manner, we have constructed a simple model similar to that of \citet{Wetzel_et_al_2013}: For an (adjustable) time $t_\text{delay}$ after accretion --- which we here define as first crossing of $r_{200,\, m}$ for consistency with their study --- galaxies are assigned the median sSFR of their matched field sample (see Section \ref{sec:onsetdetermination}). The sSFR then declines to $10^{-12}$ yr$^{-1}$ over a short period of 500 Myr, and remains at this value afterwards. Fig.~\ref{fig:quench.delaymodel} shows the resulting star-forming fraction of galaxies of stellar mass $9.5 < \log \mstar / \msun < 10.0 $ within $r_{200,\,m}$ at $z \leq 0.2$ (solid lines; the difference between different host halo masses is due to different accretion histories) as well as the values actually found in the \gimic{} simulations (dashed horizontal lines). As before, different colours represent galaxies in differently massive hosts. The intersection of each pair of lines then gives the best-fit delay time, shown by filled circles and vertical dotted lines. 

For the massive cluster (red), the inferred delay time is only $\sim 0.8$ Gyr. For less massive hosts, however, the delay time increases systematically, from $\sim 1.5$ Gyr in the case of lower-mass clusters (green) to $\sim 4$ Gyr in low-mass groups (black). While this is broadly consistent with \citet{Wetzel_et_al_2013}, there are also differences in detail. For example, these authors do not find a strong trend in $t_\text{delay}$ with host mass (see their Fig.~8). This difference could be due to shortcomings in our simulations. Alternatively, the relatively large observational uncertainties in host mass estimates may act to smear out such a strong trend in the data. Furthermore, interloping galaxies, i.e.~those which are at a three-dimensional distance $r > r_{200,\,m}$, but appear to lie at a projected radius $R < R_{200,\,m}$ may bias observational results, because they would be preferentially star-forming, and thus increase the required delay time. In our simulation-based analysis, on the other hand, we are able to cleanly separate out such interlopers. 

\subsection{Influence of pre-processed galaxies}
\begin{figure}
  \centering
    \includegraphics[width=\columnwidth]{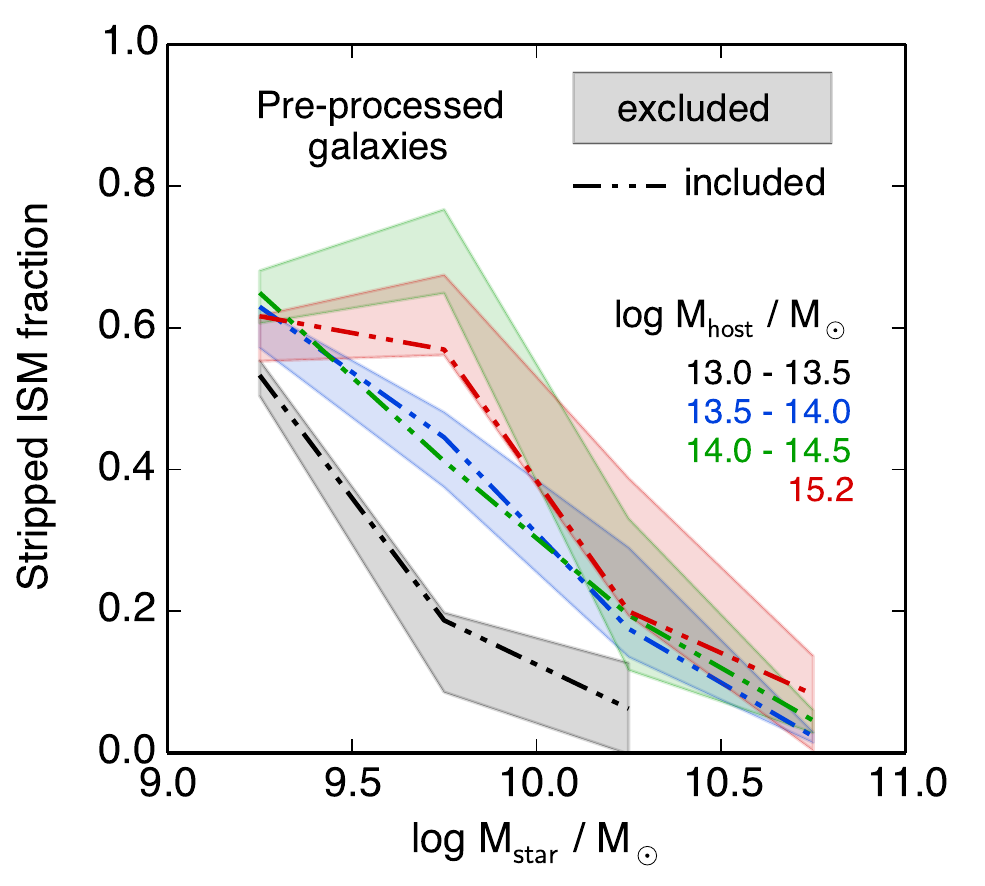}
       \caption{Effect of including pre-processed galaxies (dash-dot-dot line), compared to our default exclusion of them (shaded bands). Shown is the fraction of ISM lost through ram pressure stripping. The difference is generally small, with only one bin ($\log \mstar /\msun = [9.5, 10.0]$ in low-mass clusters) showing a significantly lower stripped fraction when pre-processed galaxies are included.}
    \label{fig:quench.prepcomp}
  \end{figure}

It is worth noting that our findings are based on a sample of galaxies explicitly selected to not have been affected by pre-processing, i.e.~the environmental effects are only due to their current host halo. In the case of group galaxies, pre-processing is expected to be rare (e.g.~\citealt{Wetzel_et_al_2013,Bahe_et_al_2013,Hou_et_al_2014}), so its inclusion cannot have a considerable effect on the results shown here. The same is not necessarily true for clusters. For completeness, we show in Fig.~\ref{fig:quench.prepcomp} a direct comparison between the stripped ISM fraction with and without inclusion of pre-processed galaxies. For most host masses, the difference is negligible, including (perhaps surprisingly) the massive cluster. Only in the lower-mass clusters (green) is there a bit more of an effect: Stripping is somewhat less strong among galaxies in the mass range $9.5 \leq \log \mstar /\msun \leq 10.0$ if those affected by pre-processing are included ($\sim 70$ vs.~$\sim 40$ per cent). This may be due to the higher fraction of quenched galaxies in this host mass bin that were pre-processed in \emph{low-mass} groups, whereas pre-processing in a massive cluster may also take place in more massive groups in which, as we have shown, stripping plays a larger role. In any case, our qualitative conclusions are unaffected: Ram pressure stripping of ISM has a significant effect on satellite galaxies, regardless of whether or not pre-processed galaxies are included.

\subsection{Applicability to real galaxies}
Are these conclusions applicable to real galaxies as well? There are two main reasons why this might not be the case: the lack of AGN feedback in \gimic{} and its SPH nature, which may suppress stripping through hydrodynamical instabilities (e.g.~\citealp{Mitchell_et_al_2009,Sijacki_et_al_2012,Zavala_et_al_2012}). The latter shortcoming may well be relevant, but would likely act to further increase the importance of stripping --- it is difficult to imagine a scenario in which these instabilities could \emph{prevent} the loss of ISM gas through stripping, noting that \citet{McCarthy_et_al_2008b} found that hydrodynamic instabilities have no significant effect on the stripping of the hot gaseous halo. The other issue, the lack of AGN feedback in \gimic{}, is potentially more subtle in its effect: it  means that massive \gimic{} galaxies are too concentrated (see \citealp{McCarthy_et_al_2012b}), and therefore presumably too resistant to stripping, but also increases their stellar mass and therefore changes their `label'. Furthermore, AGN feedback also reduces the density of the group/cluster hosts and therefore the ram pressure experienced by satellite galaxies \citep{McCarthy_et_al_2010}. However, this is expected to be most severe in the core region; e.g.~\citet{Duffy_et_al_2010} and \citet{LeBrun_et_al_2014} find that the DM and gas density, respectively, outside $\sim 0.2 \,r_{200}$, are largely unaffected by AGN feedback. The majority of stripping in \gimic{} occurs outside this central region (Fig.~\ref{fig:quench.rpprofile}), which suggests that the ram pressure experienced by them is modelled accurately. 

A further shortcoming in \gimic{} that is likely related to the absence of AGN feedback is the lack of a `mass quenching' effect, by which more massive galaxies are less likely to form stars \citep{Peng_et_al_2010}. This is obviously regrettable in terms of producing a realistic galaxy population that can be quantitatively compared to observations, but on the positive side, it means that environmental influence is the \emph{only} physical effect quenching \gimic{} galaxies. Furthermore, although the \gimic{} simulations contain an unrealistically high abundance of star-forming massive galaxies, the properties of the \emph{individual} objects (including their star formation rates), agree well with observations \citep{McCarthy_et_al_2012b}. The insight gained from studying the transformation of our simulated galaxies is therefore most likely applicable to real ones as well, at least at a qualitative level.

The best way to deduce the influence of the above-mentioned deficiencies of the \gimic{} suite would be to repeat this analysis with an improved set of simulations. A promising new tool in this regard, the \textsc{Eagle} simulations, has recently been introduced by \citet{Schaye_et_al_2014}. Their higher mass and time resolution, as well as significantly improved sub-grid physics (i.e.~inclusion of AGN feedback) and hydrodynamics implementations, will help to better understand the complex details of how group and cluster galaxies transition from star forming blue spirals to red, passive ellipticals. 

\section*{Acknowledgments}
We are grateful to the referee for their many helpful suggestions, and the members of the \gimic{} team for allowing us to analyse the simulations. We thank Michael Balogh, Marcella Carollo, Guinevere Kauffmann, Sean McGee, Thorsten Naab, Tom Theuns and Simon White for helpful discussions. YMB acknowledges support from a postgraduate award from STFC and an MPA fellowship, and IGM from a STFC Advanced Fellowship.  The \textsc{gimic} simulations were carried out using the HPCx facility at the Edinburgh Parallel Computing Centre (EPCC) as part of the EC's DEISA `Extreme Computing Initiative' and with the Cosmology Machine at the Institute of Computational Cosmology of Durham University. Several figures in this paper were produced using the \texttt{matplotlib} \citep{Hunter_2007} and \texttt{astropy} \citep{Astropy_2013} packages.

\bibliographystyle{mn2e}
\bibliography{thesis}


\begin{appendix}

\section{Galaxy tracing}
\label{sec:app_tracing}

In this Appendix, we describe in detail our method to trace subhaloes between different snapshots in our simulations. This procedure is based on the ability to identify individual particles in subsequent snapshots through their unique IDs. We link in each pair of neighbouring snapshots any two subhaloes that share at least 20 dark matter or star particles, or fewer if their total mass accounts for more than 50 per cent of the total mass of at least one of the subhaloes. In the simplest situation, there is only one such link going from (`sent') and leading to (`received by') each subhalo: in this case, we could unambiguously identify the same subhaloes in both snapshots.

In reality, however, subhaloes may exchange particles between each other (e.g.~in mergers), so that one subhalo identified in snapshot $i$ has, in general, no identical counterpart in a second snapshot $j$. This leads to one subhalo in $i$ being linked to multiple others in $j$ (and vice versa), so that we have to find the `closest match' out of these. For this purpose we rank all links sent from or received by the same subhalo by their total mass, which gives each link a sender and receiver `rank'. We then `select' links (i.e.~identify those which connect the ``same'' subhalo in two snapshots) in order of these ranks until either all subhaloes are connected, or more realistically, until no more links are available. In case of a conflict between two links with the same ranks in inverted order (e.g.~sender rank 1, receiver rank 2 and sender rank 2, receiver rank 1), we select the one with higher \emph{receiver} rank, i.e.~the one contributing a larger fraction of mass to its descendant galaxy, rather than the one carrying most mass away from its progenitor. The reason for this choice is that it prevents situations where a small subhalo accreted onto a bigger subhalo is misidentified as the bigger subhalo's descendent, while allowing subhaloes that lose the majority of their mass to a bigger system (e.g.~a cluster halo) to be traced for as long as possible. We repeat this process for each pair of neighbouring snapshots to obtain a continuous history of all subhaloes in our simulation.

An additional complication is that subhalo finders such as \textsc{subfind} have difficulty identifying subhaloes in dense backgrounds, such as the central regions of a galaxy cluster (e.g.~\citealt{Muldrew_et_al_2011}). Unaccounted for, this would lead to spurious subhalo ``disruption'' (when a subhalo is still physically existing, but not identified as such) and ``formation'' (if it is re-identified later). To mitigate this, we also trace subhaloes over 2 snapshot intervals by forming what we call `long links' between each pair of snapshots separated by one snapshot between them, in analogy to the `short links' described above. In the simplest case, the temporary non-identification will leave a subhalo A in the first snapshot $i$ without a (short-link) descendant, and a counterpart B in the second snapshot ($k$) without a (short-link) progenitor. Provided A and B are connected by a long link, we can then join them together from $i$ to $k$ and skip the missing identification in snapshot $j$ in-between.  

However, it is also possible that between redshifts $i$ and $k$, the subhalo accretes another, smaller subhalo, which would then be identified as its progenitor although physically it is not (c.f.~above). We therefore also allow selection of long links between subhaloes that already have an identified (short-link) progenitor or descendant in the immediately neighbouring snapshot, provided this results in a better match of particles between subhaloes. This allows our method to robustly follow self-bound structures through time, accounting for subhalo formation, merging and disruption, as well as temporary non-identification of subhaloes in dense environments.

\section{Reason for ram pressure scatter}
\label{sec:app.rpscatter}
\begin{figure}
  \centering
    \includegraphics[width=\columnwidth]{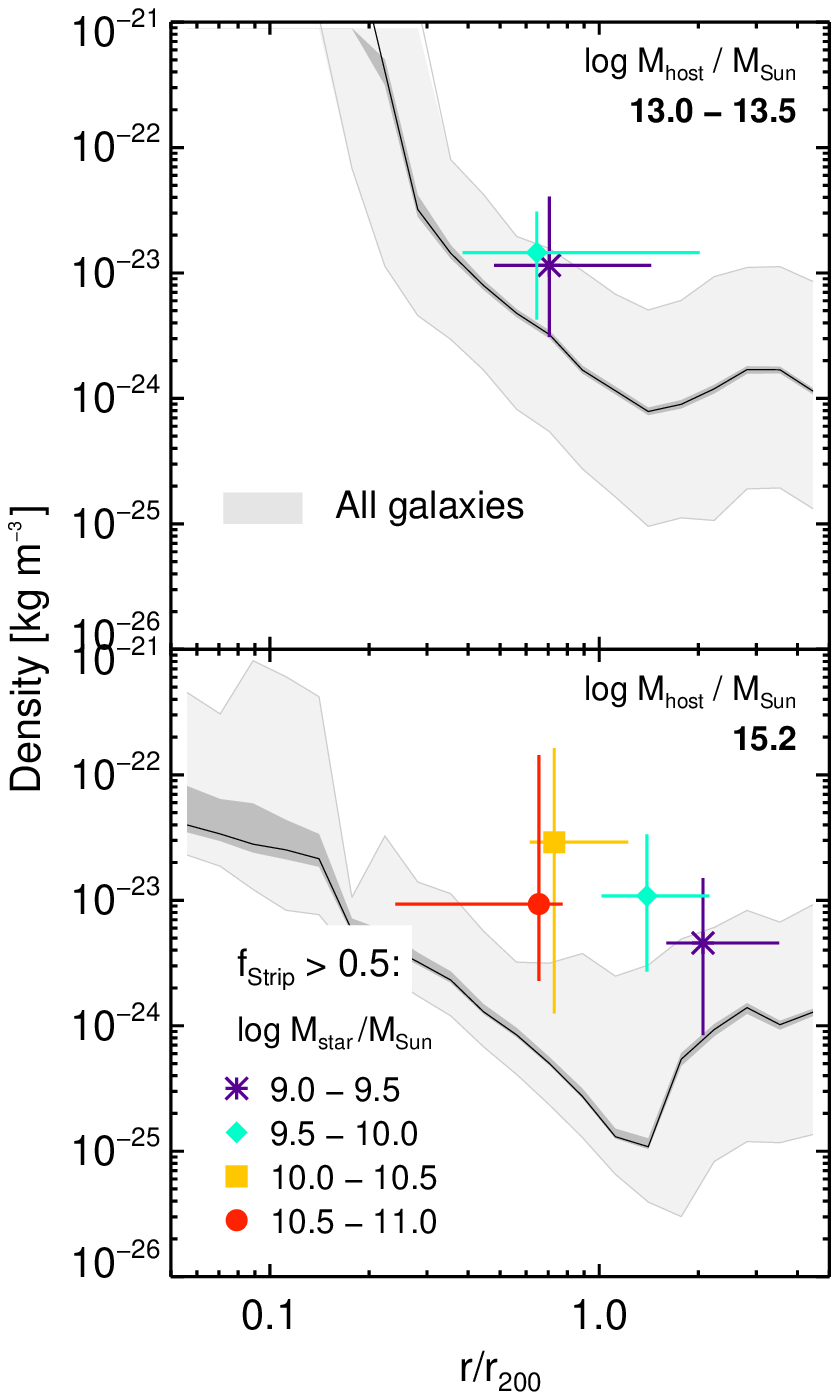}
       \caption{Radial density profiles for low mass groups (top) and the massive cluster (bottom), in analogy to the ram pressure profiles in Fig.~\ref{fig:quench.rpprofile}.}
    \label{fig:quench.rhoprofile}
  \end{figure}

\begin{figure}
  \centering
    \includegraphics[width=\columnwidth]{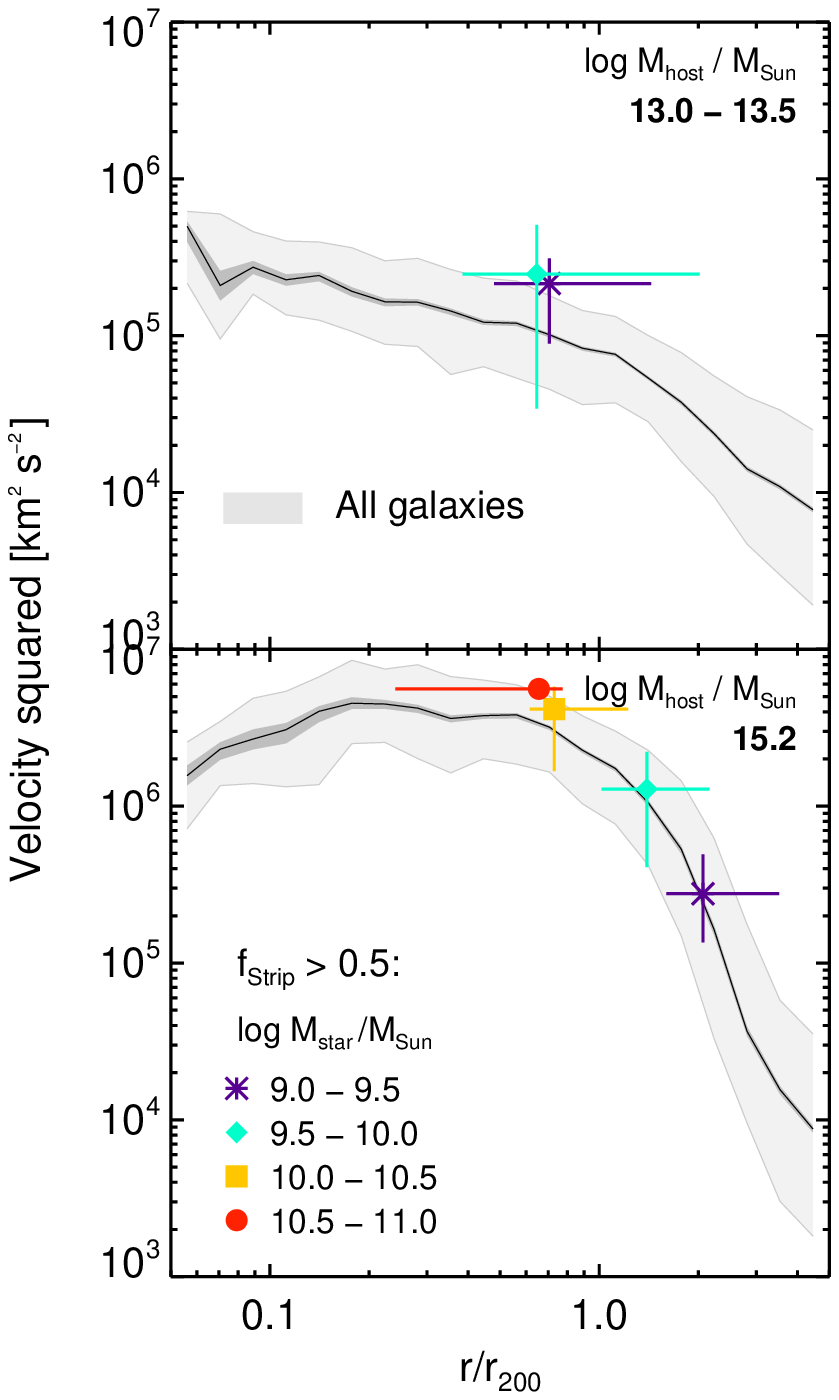}
       \caption{Radial velocity profiles for low mass groups (top) and the massive cluster (bottom), in analogy to the ram pressure profiles in Fig.~\ref{fig:quench.rpprofile}. Note that this shows velocity squared for a direct comparison with the density profile in Fig.~\ref{fig:quench.rhoprofile}.}
    \label{fig:quench.vsqprofile}
  \end{figure}

In Section \ref{sec:competition} we had shown that the ram pressure at a given group- or cluster-centric radius $r/r_{200}$ can vary by several orders of magnitude, and that galaxies actually being stripped are preferentially exposed to atypically high levels of ram pressure. In this Appendix, we analyse the origin of this ram pressure scatter. To this end, we show in Figs.~\ref{fig:quench.rhoprofile} and \ref{fig:quench.vsqprofile} the radial profiles of ICM density $\rho$ and ICM velocity (squared) relative to the galaxy, $|\mathbf{v}|^2$, respectively. For the low-mass groups (top), stripped galaxies are typically surrounded by ICM which is both slightly denser and moving faster with respect to them than the radial average, with both factors contributing roughly equally to the positive bias in ram pressure. The situation is very different for the massive cluster, however. Here, the ICM velocity relative to stripped galaxies is very close to the spherical average (black line), with only a hint of a positive offset for massive galaxies (red/yellow) which are being stripped preferentially at $r \leq r_{200}$. In contrast, the ICM density surrounding these galaxies exceeds the average by up to two orders of magnitude. We can therefore conclude that, in the case of galaxy clusters, ram pressure stripping is largely restricted to \emph{overdense} regions, in agreement with the independent results of \citet{Tonnesen_Bryan_2008}. In \citet{Bahe_et_al_2013}, we had shown that these overdense regions in clusters correspond to filaments. Our results here therefore imply that these filaments not only cause enhanced stripping of hot gas at radii far beyond $r_{200}$, but also of the star-forming ISM at smaller radii. 

We finally note that, in contrast to the ram pressure profiles in Fig.~\ref{fig:quench.rpprofile}, both the density and velocity profiles \emph{do} show a break around $r = r_{200}$, especially in the case of the massive cluster (bottom panels). Outside this radius, the ICM density around galaxies increases with radius --- a consequence of their preferential location inside overdense filaments --- whereas the ICM velocity declines rather sharply, due to the increased degree of co-flow between galaxies and the surrounding gas (see \citealp{Bahe_et_al_2013}). Multiplied together, these two effects largely cancel out which causes the smooth ram pressure profile either side of $r_{200}$.

\end{appendix}
\end{document}